\documentclass{article}

     \usepackage[final]{neurips_2025}

\usepackage[utf8]{inputenc} 
\usepackage[T1]{fontenc}    
\usepackage{hyperref}       
\usepackage{url}            
\usepackage{booktabs}       
\usepackage{amsfonts}       
\usepackage{nicefrac}       
\usepackage{microtype}      
\usepackage[table]{xcolor}
\usepackage{amssymb}

\usepackage{sidecap}

\usepackage{enumitem}
\usepackage{amsmath}
\usepackage{multirow}
\usepackage{graphicx}
\usepackage{wrapfig}
\newcommand{\figtechniRD}{
\begin{figure}[t]
    \centering
    \includegraphics[width=0.95\textwidth]{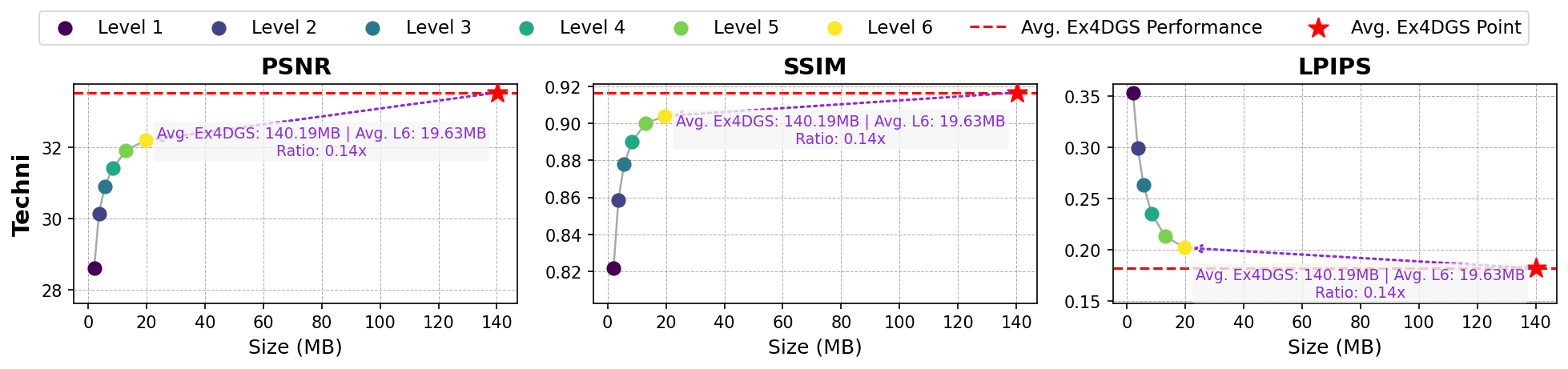}
    \vspace{-2mm}
    \caption{\textbf{Average RD performance comparison between our method (Levels 1-6) and the Ex4DGS~\cite{ex4dgs} on the Technicolor dataset.} Per-scene details are provided in \Cref{appendix:quantitative}.}
    \label{fig:rdo_techni}
\vspace{-5mm}
\end{figure}
}

\newcommand{\figntvRD}{
\begin{figure}[t]
    \centering
    \includegraphics[width=0.95\textwidth]{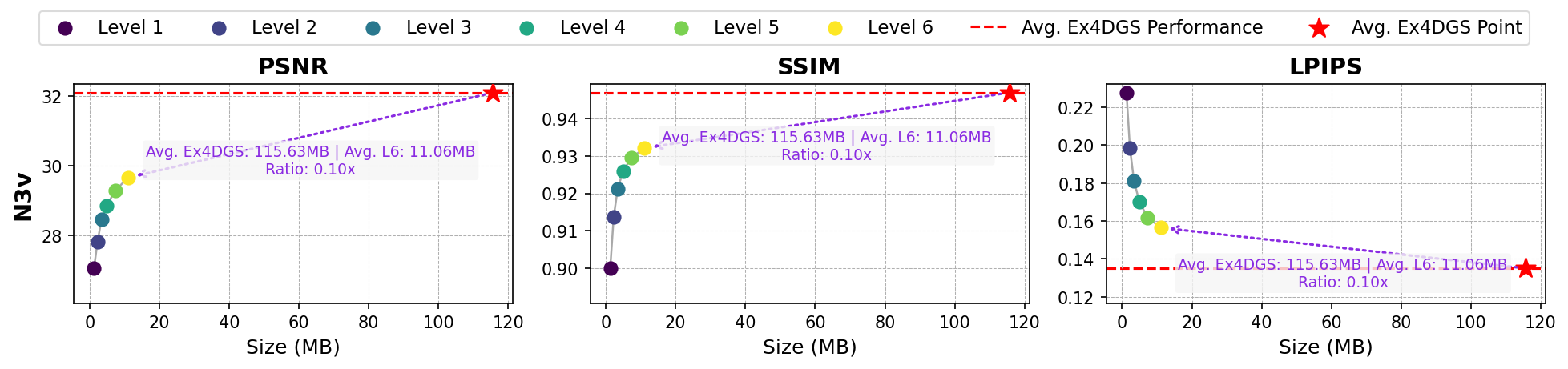}
    \vspace{-2mm}
    \caption{\textbf{Average RD performance comparison between our method (Levels 1-6) and the Ex4DGS~\cite{ex4dgs} on the Neural 3D Video~(N3V) dataset.} Per-scene details are provided in \Cref{appendix:quantitative}.}
    \label{fig:rdo_n3v}
\vspace{-3mm}
\end{figure}
}

\newcommand{\figtechniRDEach}{
\begin{figure}[h!]
    \centering
    \includegraphics[width=0.95\textwidth]{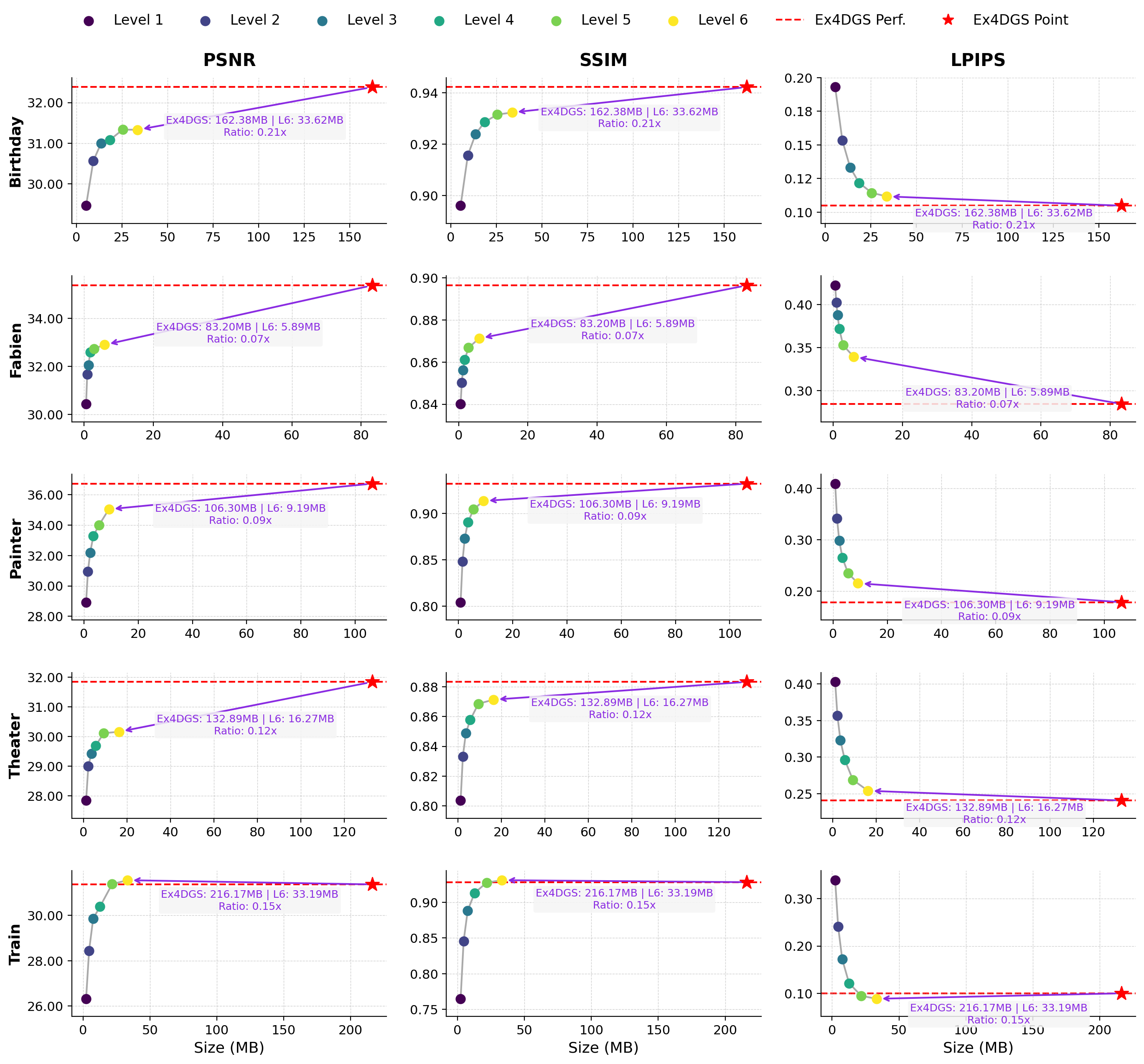}
    \vspace{-2mm}
    \caption{\textbf{Per-scene RD performance comparison between our method (Levels 1-6) and the Ex4DGS~\cite{ex4dgs} on the Technicolor dataset.}}
    \label{fig:rdo_techni_perscene}
\end{figure}
}

\newcommand{\figntvRDEach}{
\begin{figure}[h!]
    \centering
    \includegraphics[width=0.95\textwidth]{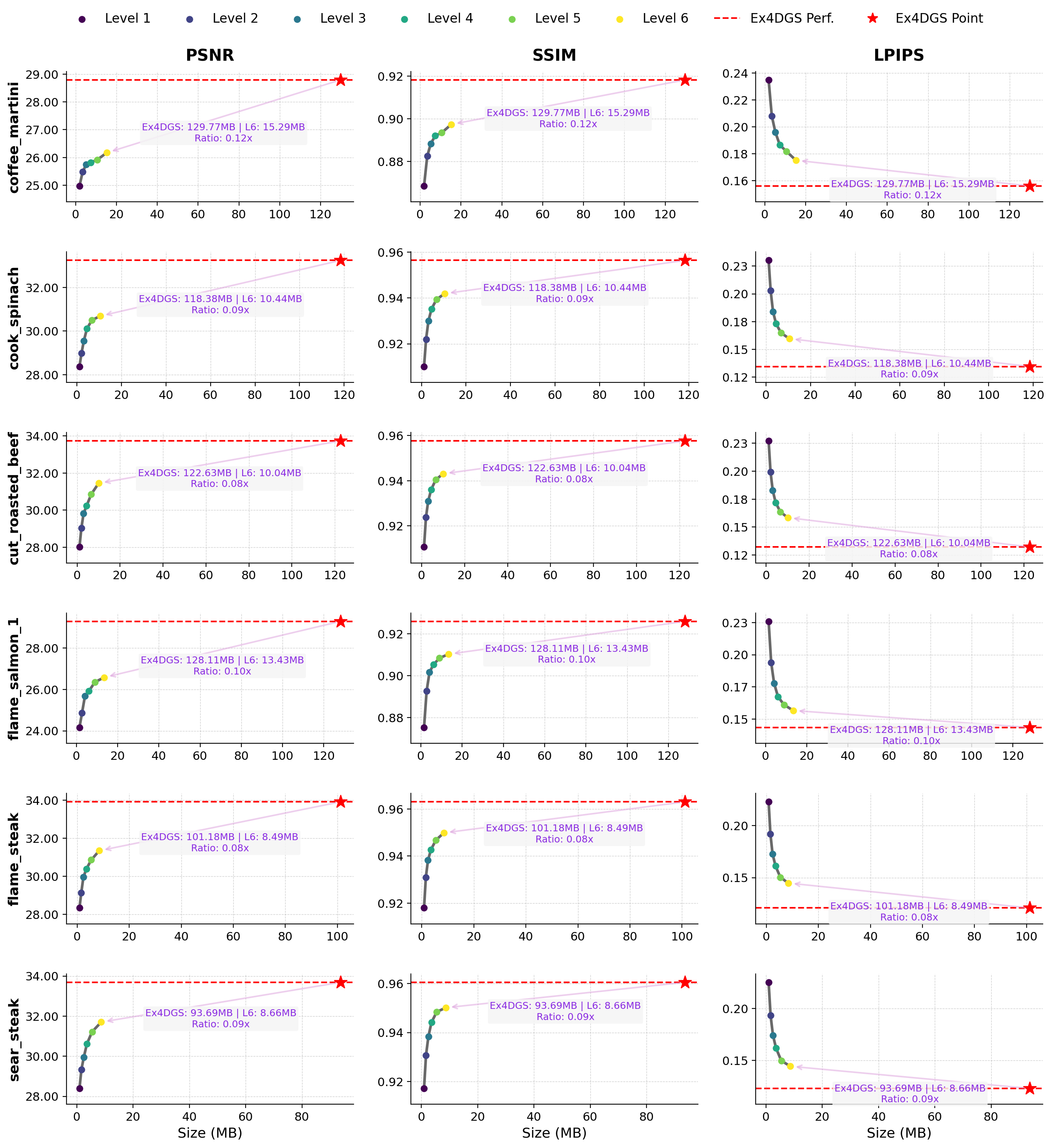}
    \vspace{-2mm}
    \caption{\textbf{Per-scene RD performance comparison between our method (Levels 1-6) and the Ex4DGS~\cite{ex4dgs} on the Neural 3D Video~(N3V) dataset.}}
    \label{fig:rdo_n3v_perscene}
\vspace{-3mm}
\end{figure}
}

\newcommand{\figtechniQual}{
\begin{figure}[t]
    \centering
    \includegraphics[width=0.95\textwidth]{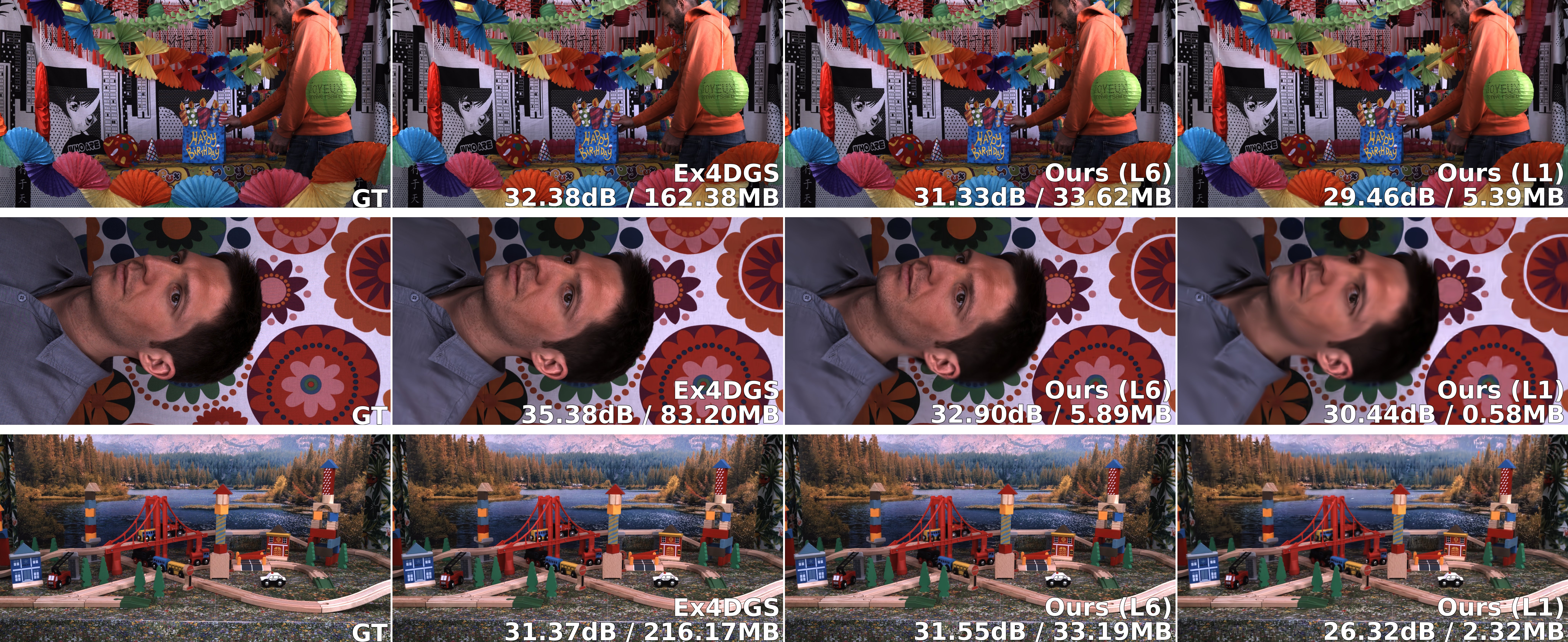}
    \caption{\textbf{Qualitative results on the Technicolor dataset.} We compare Ground Truth, Ex4DGS~\cite{ex4dgs}, and ours at compression levels 6 and 1. PSNR (dB) / Size (MB) are shown below each image.}
\vspace{-5mm}
\label{fig:qual_techni}
\end{figure}
}

\newcommand{\fignthreevQual}{
\begin{figure}[t]
    \centering
    \includegraphics[width=0.85\textwidth]{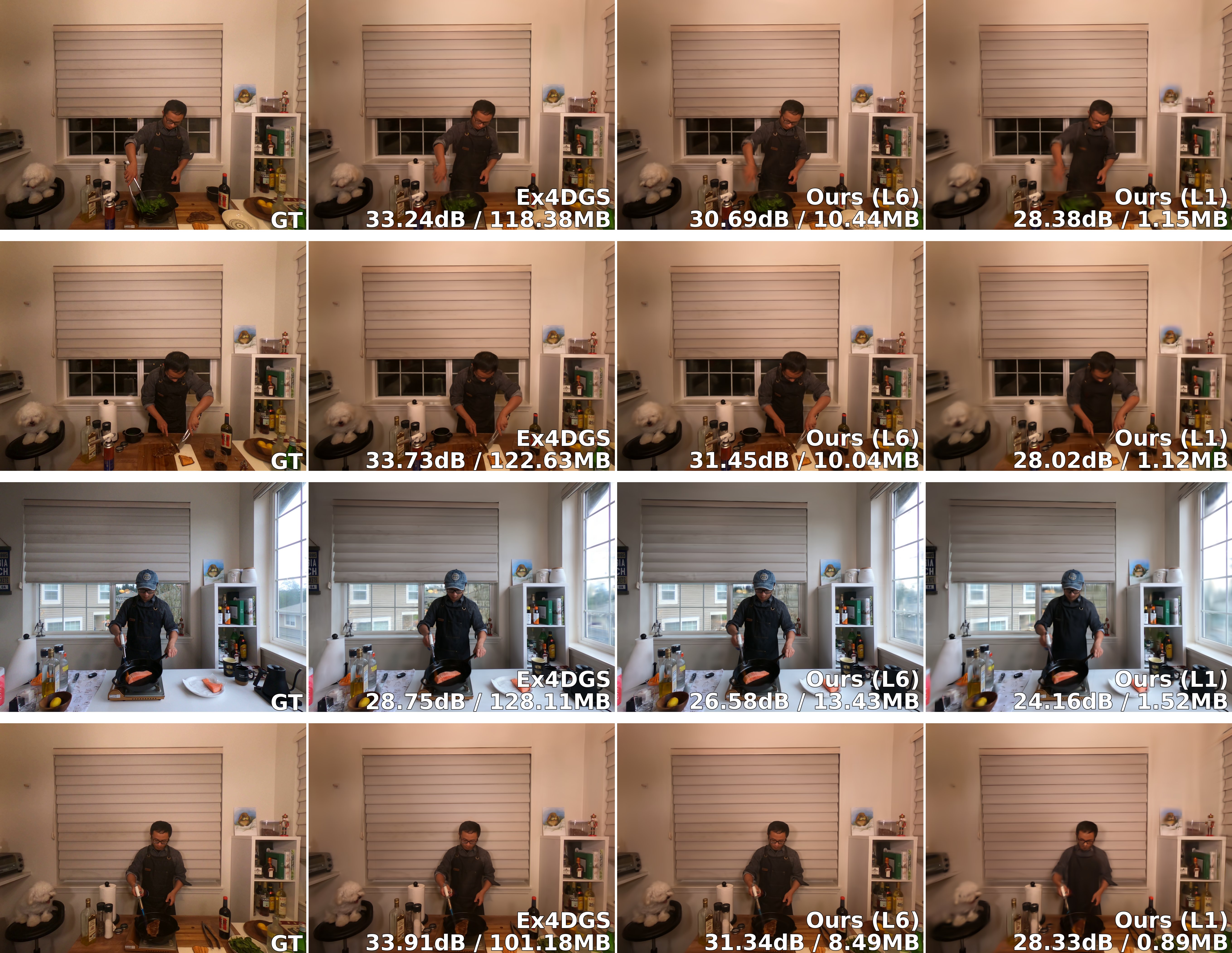}
    \caption{\textbf{Qualitative results on the N3V dataset.} We compare Ground Truth, Ex4DGS~\cite{ex4dgs}, and ours at compression levels 6 and 1. PSNR (dB) / Size (MB) are shown below each image.}
\vspace{-5mm}
\label{fig:qual_n3v}
\end{figure}
}

\newcommand{\figntvablation}{
\begin{SCfigure}[1.0][t] 
    \centering 
    \includegraphics[width=0.6\linewidth]{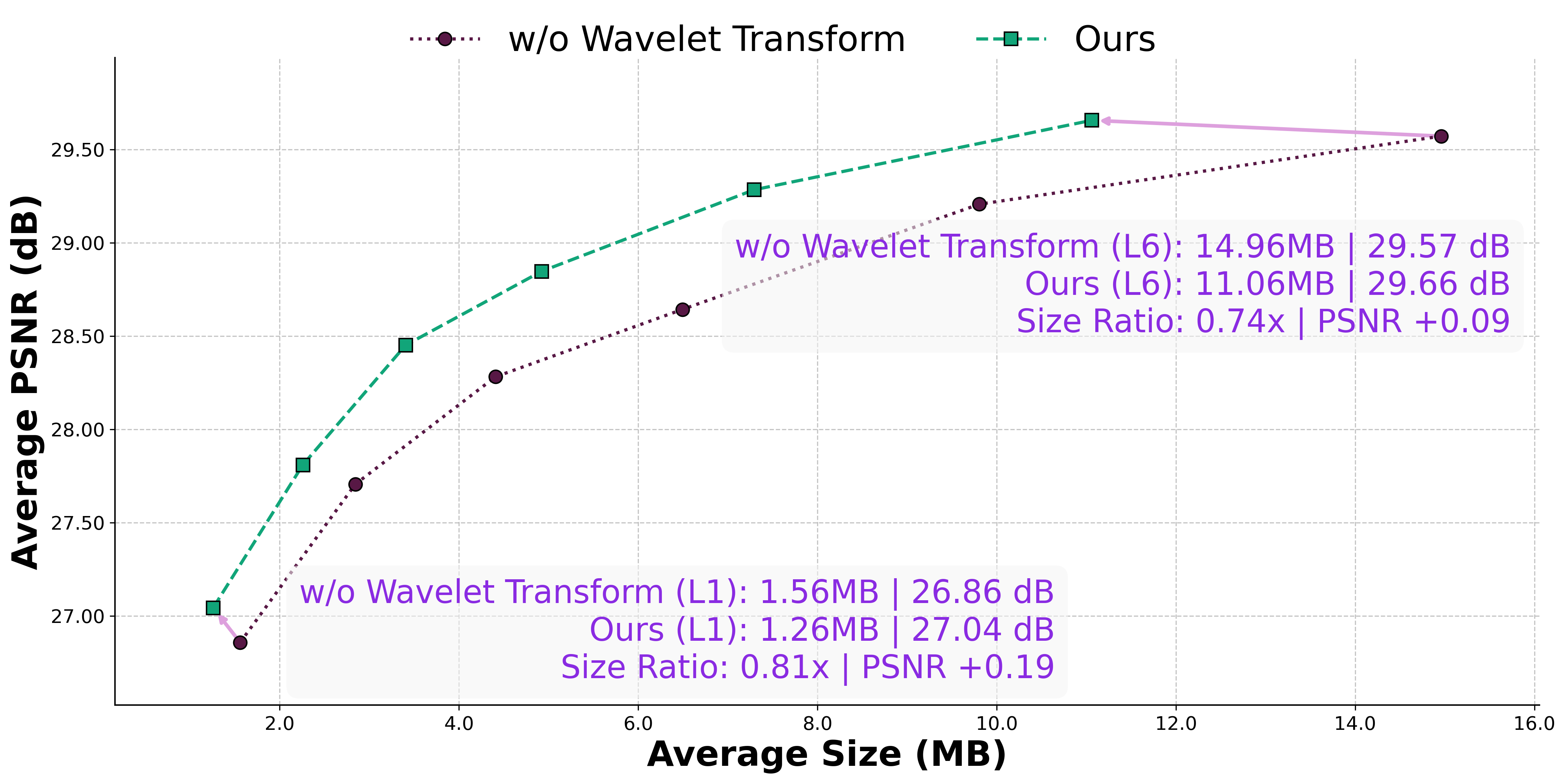} 
    \caption{\textbf{Rate-distortion impact of using wavelet transform for dynamic positions on the N3V dataset.} We compare the proposed method (`Ours', with wavelet transform) against the same model without wavelet transform (`w/o Wavelet Transform'). Axes show avg. PSNR (dB) vs. avg. model size (MB).}
    \label{fig:ablation}
\vspace{-4mm}
\end{SCfigure}
}

\newcommand{\figStaticDynamicTechni}{
\begin{figure}[!h]
    \centering
    \includegraphics[width=0.9\linewidth]{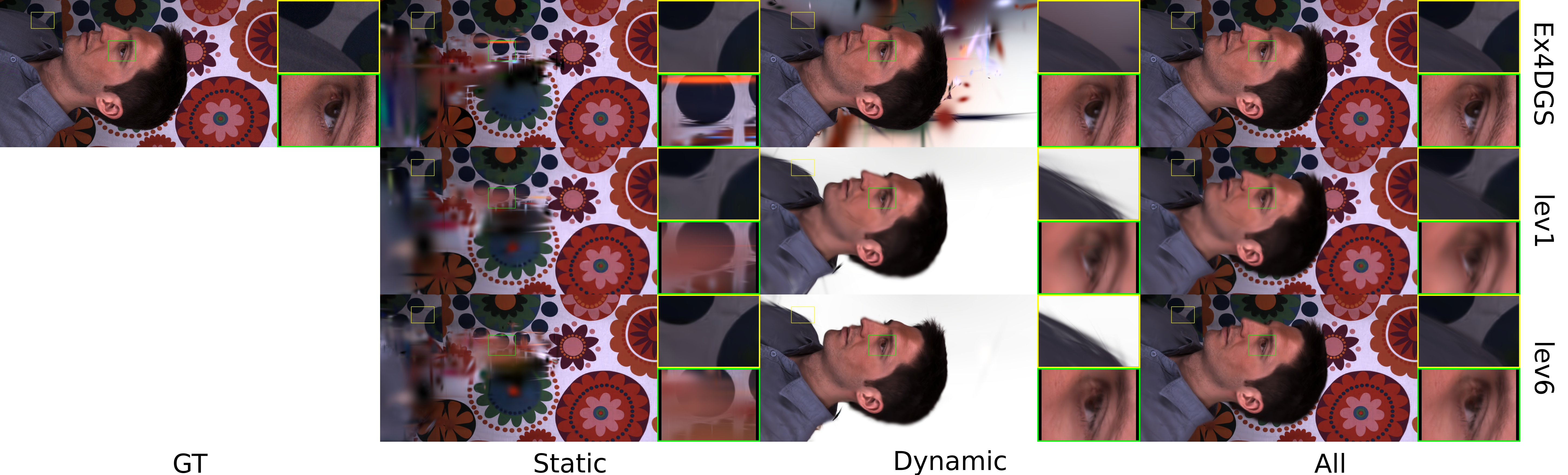}
    \includegraphics[width=0.9\linewidth]{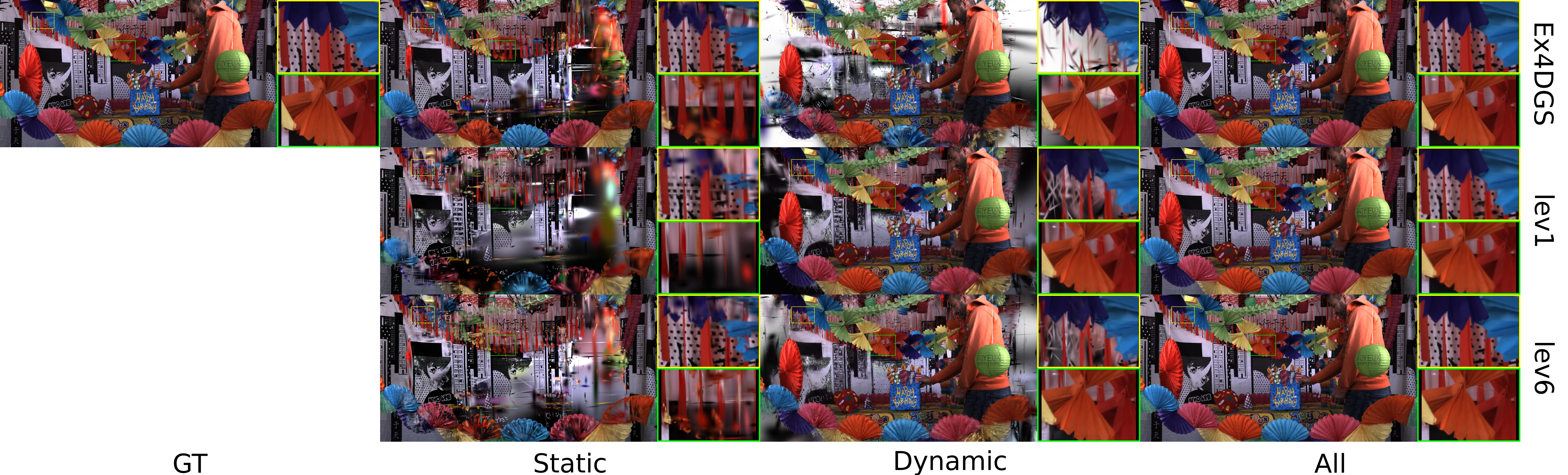}
    \includegraphics[width=0.9\linewidth]{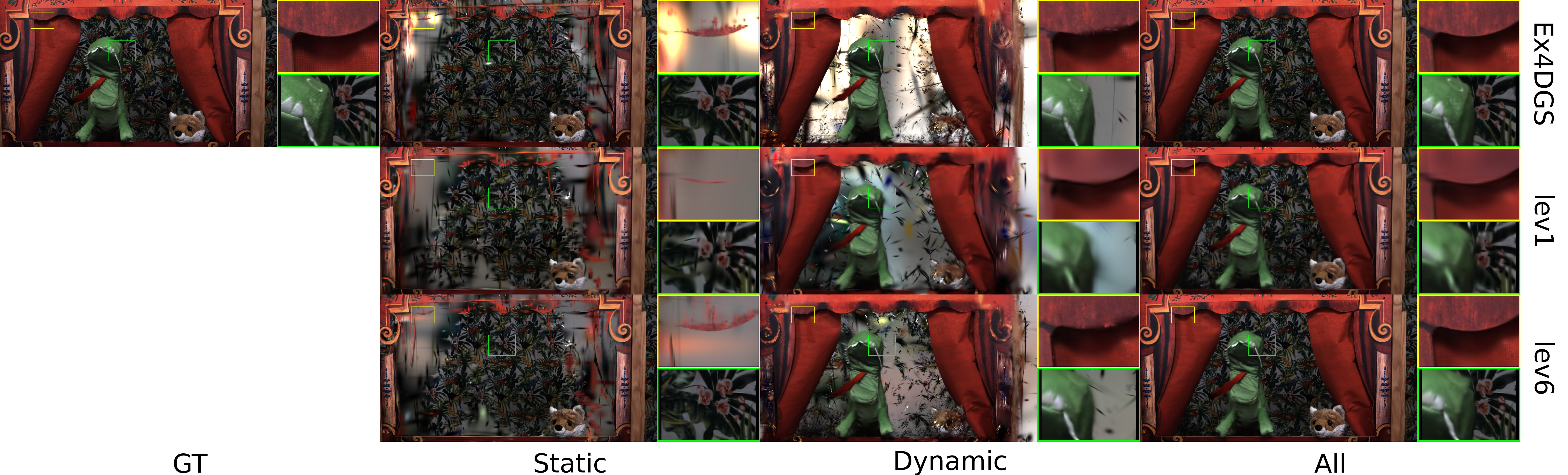}
    \includegraphics[width=0.9\linewidth]{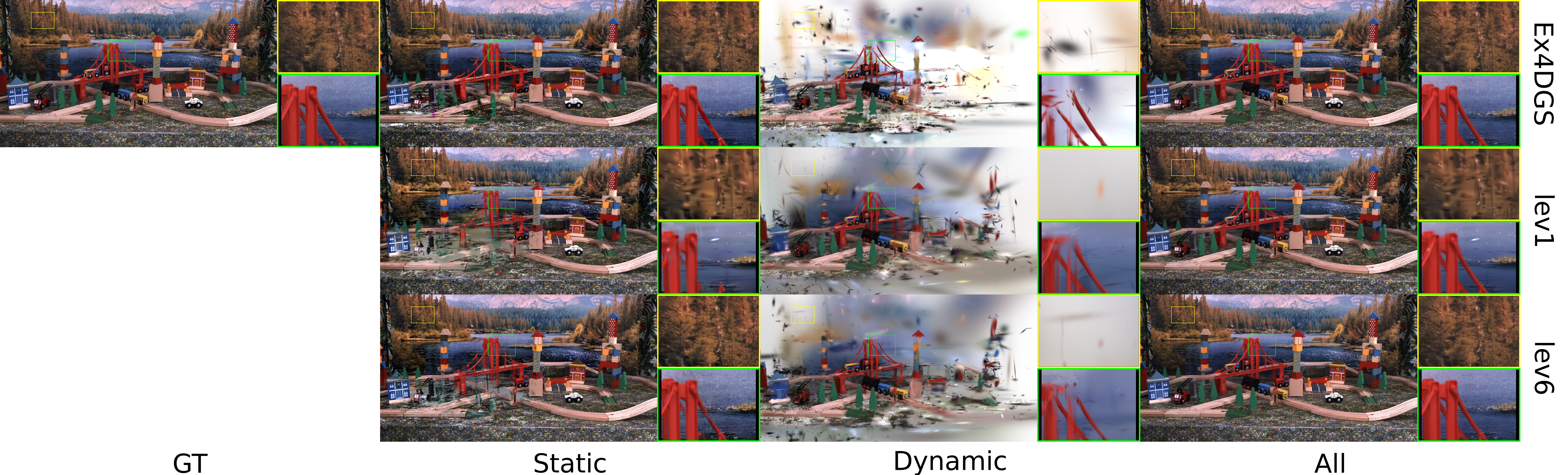}
    \includegraphics[width=0.9\linewidth]{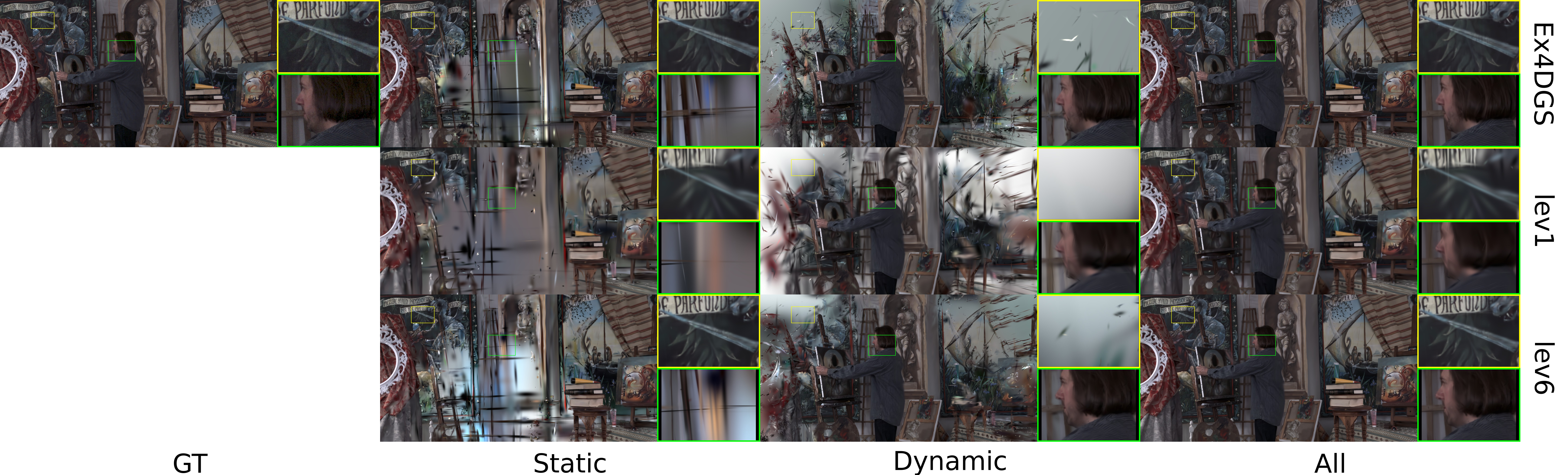}
    \caption{\textbf{Visualization of static and dynamic points on Technicolor dataset.} From top to bottom: Fabien, Birthday, Theater, Train, and Painter.}
    \label{fig:staticdynamic}
\end{figure}
}

\newcommand{\figStaticDynamicNThreeVFirst}{
\begin{figure}[!h]
    \centering
    \includegraphics[width=0.9\linewidth]{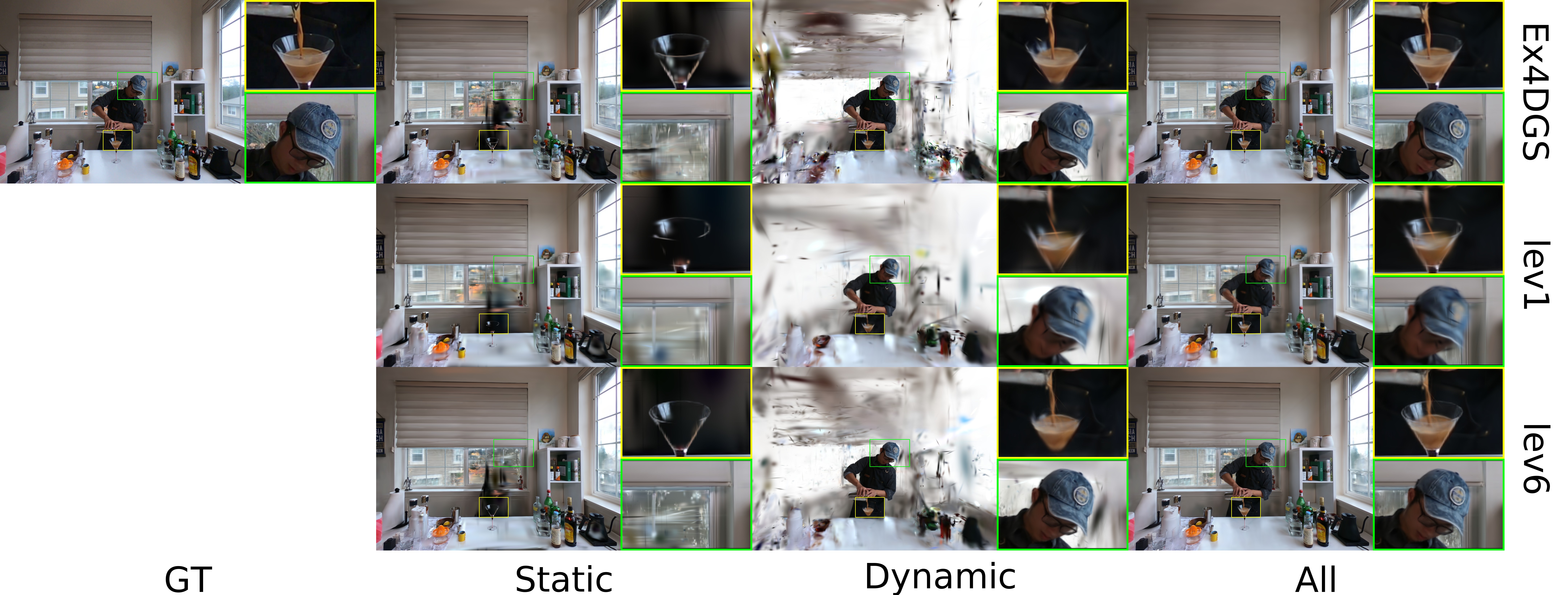}
    \includegraphics[width=0.9\linewidth]{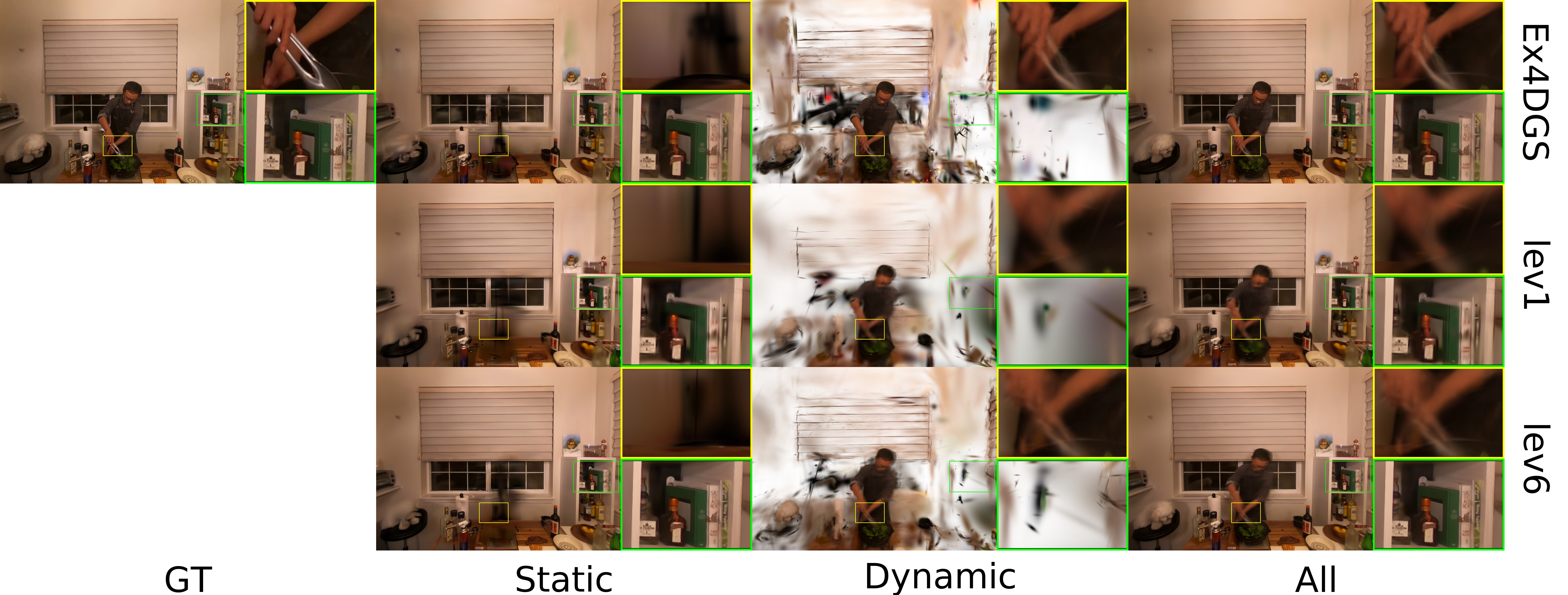}
    \includegraphics[width=0.9\linewidth]{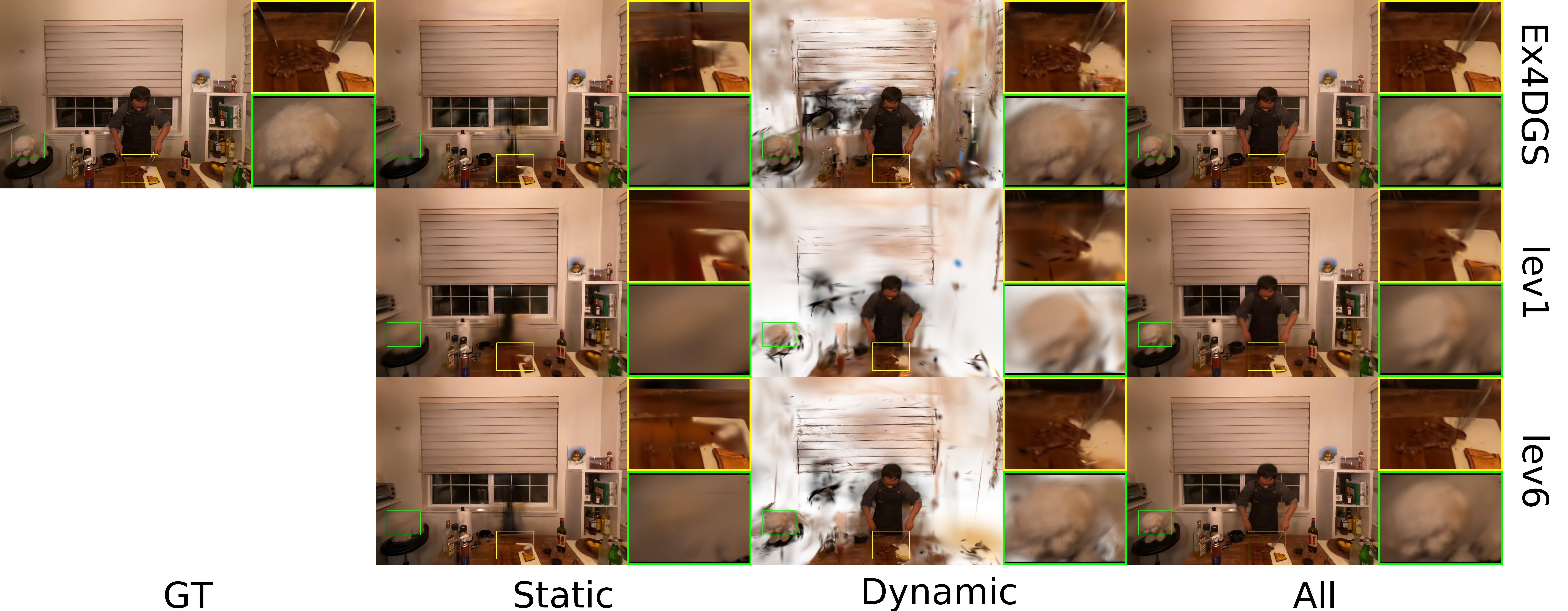}
    \caption{\textbf{Visualization of static and dynamic points on N3V dataset.} From top to bottom: Coffee Martini, Cook Spinach, and Cut Roasted Beef.}
    \label{fig:staticdynamic_nthreev_first}
\end{figure}
}

\newcommand{\figStaticDynamicNThreeVSecond}{
\begin{figure}[!h]
    \centering
    \includegraphics[width=0.9\linewidth]{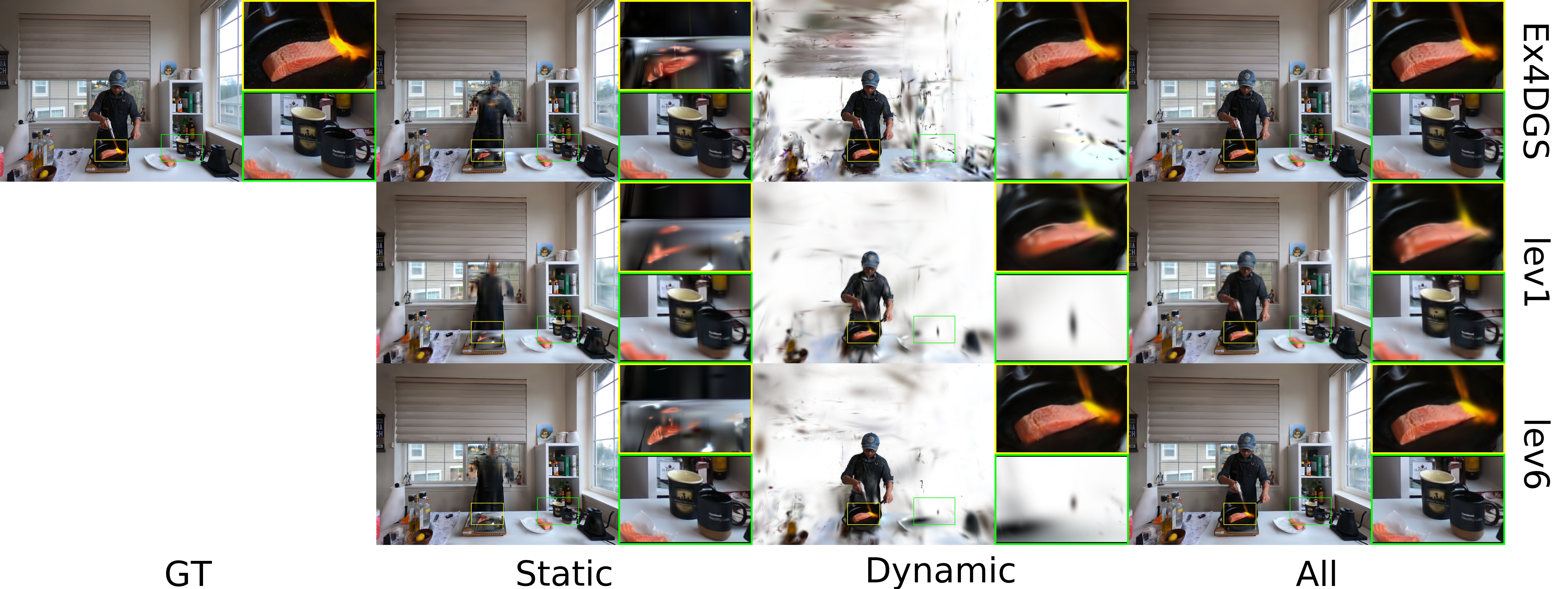}
    \includegraphics[width=0.9\linewidth]{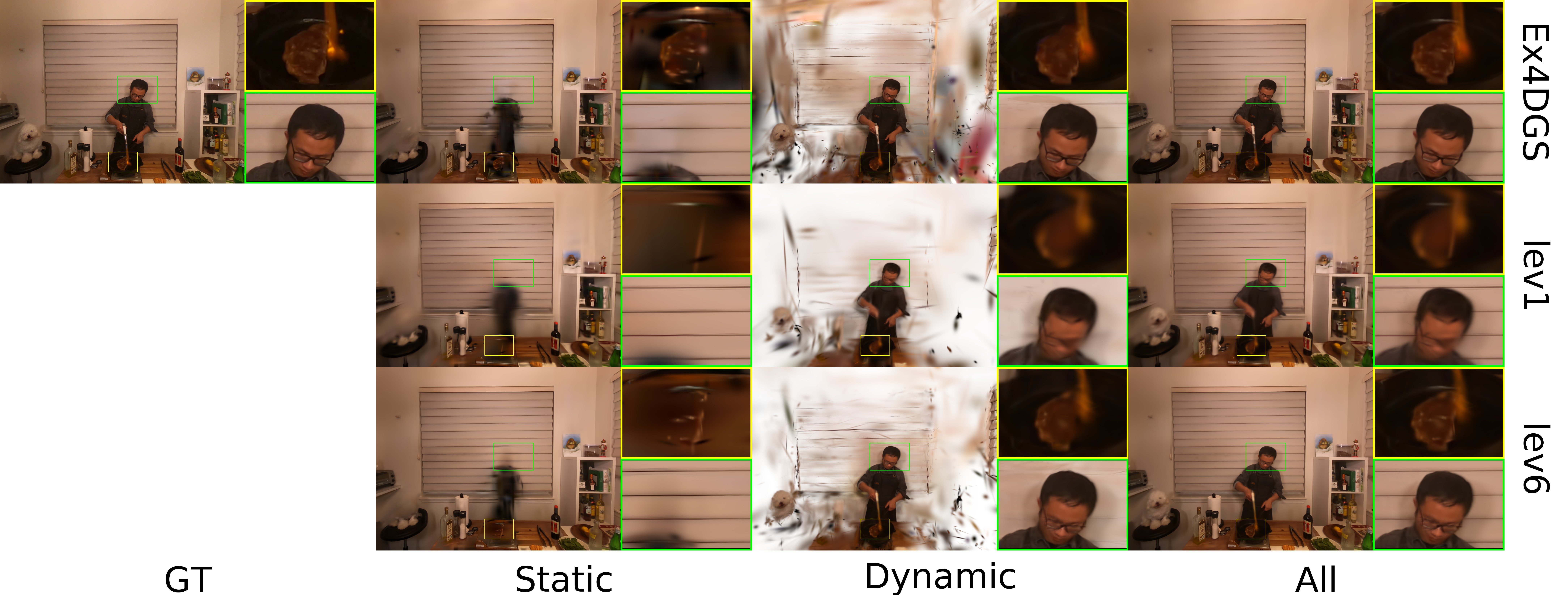}
    \includegraphics[width=0.9\linewidth]{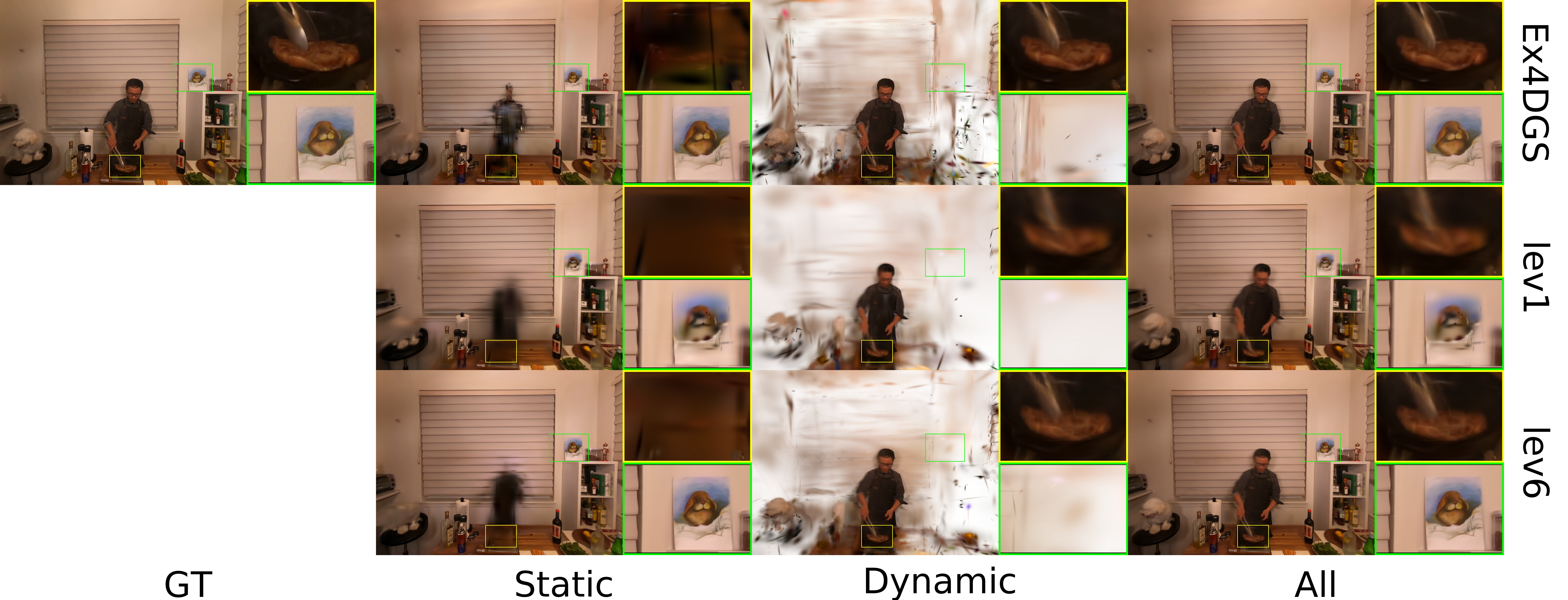}
    \caption{\textbf{Visualization of static and dynamic points on N3V dataset.} From top to bottom: Flame Salmon, Flame Steak, and Sear Steak.}
    \label{fig:staticdynamic_nthreev_second}
\end{figure}
}

\newcommand{\tworow}[2]{\begin{tabular}[c]{@{}c@{}}#1\vspace{-2pt}\\#2\end{tabular}}

\newcommand{\tabNthreeV}{
\begin{table}[t]
\centering
\caption{\textbf{Quantitative results on the N3V dataset comparing PSNR (dB), average model size (MB), and FPS.} Our proposed RD-optimized method offers flexible compression levels. Level 6 achieves a significant size reduction (e.g., about 10$\times$ reduction to 11.06MB) compared to the Ex4DGS~\cite{ex4dgs} while maintaining competitive quality. Level 1 demonstrates the capability for aggressive compression (about 90$\times$ size reduction to 1.26MB) for highly resource-constrained scenarios.}
\vspace{-2mm}
\resizebox{\linewidth}{!}{

\begin{tabular}{cccccccccc}
\hline
\toprule
\multirow{2}{*}{Model\vspace{-15pt}} & \multicolumn{7}{c}{PSNR (dB)$\uparrow$} & ~~~~MB$\downarrow$~~~~ & Frame/s$\uparrow$ \\ \cmidrule(lr){2-8} \cmidrule(lr){9-9} \cmidrule(lr){10-10} 
 & \tworow{Coffee}{Martini} & \tworow{Cook}{Spinach} & \tworow{\!\!\!\!\!Cut\,Roasted\!\!\!\!\!}{Beef} & \tworow{Flame}{Salmon} & \tworow{Flame}{~Steak~} & \tworow{Sear}{~Steak~} & Average & \tworow{Size}{(Average)} & FPS\\ \midrule
\multicolumn{10}{c}{NeRF-based}   \\
NeRFPlayer~\cite{song2023nerfplayer}   & 31.53 & 30.56 & 29.35 & 31.65 & 31.93 & 29.13 & 30.69 & 5130 & 0.05 \\ 
HyperReel~\cite{attal2023hyperreel} & 28.37& 32.30& 32.92& 28.26& 32.20& 32.57& 31.10& 360& 2 \\ 
K-Plane~\cite{fridovich2023k} & 29.99 & 32.60 & 31.82 & 30.44 & 32.38 & 32.52 & 31.63 & 311 & 0.3  \\
MixVoxels-L~\cite{wang2023mixed} & 29.63 & 32.25 & 32.40 & 29.81 & 31.83 & 32.10 & 31.34 & 500 & 37.7 \\
MixVoxels-X~\cite{wang2023mixed} & 30.39 & 32.31 & 32.63 & 30.60 & 32.10 & 32.33 & 31.73 & 500 & 4.6 \\

\midrule
\multicolumn{10}{c}{Gaussian Splatting-based}   \\
4DGS~\cite{4dgs}   & 26.51 & 32.11 & 31.74 & 26.93 & 31.44 & 32.42 & 30.19 & 6057 & 72.0\\
4DGaussians~\cite{yang2023real}   &  26.69 & 31.89 & 25.88 & 27.54 & 28.07 & 31.73 & 28.63 & 34 & 146.6 \\
3DGStream~\cite{sun20243dgstream}   &  27.75 & 33.31 & 33.21 & 28.42 & 34.30 & 33.01 & 31.67 & 1200 & -\\
Ex4DGS~\cite{ex4dgs} & 28.79       & 33.23     & 33.73 & 29.29     & 33.91    & 33.69   & 32.11   & 115  & 72.3 \\ 
\hline
Ours [$\level{L}$1]   &  24.97 & 28.38 & 28.02 & 24.16 & 28.33 & 28.39 & 27.04 & 1.26 \scriptsize{\textcolor{red}{(-98.9\%)}} & 163.0  \\
Ours [$\level{L}$6]   &  26.17 & 30.69 & 31.45 & 26.58 & 31.34 & 31.70 & 29.66 & 11.06 \scriptsize{\textcolor{red}{(-90.4\%)}} & 100.9  \\
\bottomrule
\end{tabular}
}
\label{tbl:n3v_compare}
\vspace{-5mm}
\end{table}
}

\newcommand{\tabTechni}{

\begin{wraptable}{r}{0.55\linewidth}
\vspace{-4mm}
\centering
\caption{\textbf{Quantitative results averaged over the Technicolor dataset.} We compare the proposed method ($\level{L}$1, $\level{L}$6) against the Ex4DGS~\cite{ex4dgs} using standard quality metrics and model size.}
\vspace{1mm}
\scriptsize
\begin{tabular}{cccccc}
\toprule
 & PSNR & SSIM & LPIPS & Size(MB) & FPS(s)\\ \midrule
Ex4DGS~\cite{ex4dgs} & 33.62 & 0.916 & 0.088 & 140.2 & 72.3\\
Ours [$\level{L}$6] & 32.20 & 0.904 &  0.100 & 19.6\tiny{\textcolor{red}{(-86.0\%)}} & 113.1\\
Ours [$\level{L}$1] & 28.60 & 0.822 &  0.232 & 2.1\tiny{\textcolor{red}{(-98.5\%)}} &  213.9\\
\bottomrule
\end{tabular}
\label{tbl:techni_compare}
\vspace{-3mm}

\end{wraptable}

}

\newcommand{\tabCompRatioSD}{

\begin{table}[h!]
\caption{\textbf{Reduction in static ($\mathcal{G}_s$) and dynamic ($\mathcal{G}_d$) point counts for Technicolor scenes.}}
\centering
\begin{tabular}{ccll|ccll}
\toprule
 & \multicolumn{1}{c}{} & \multicolumn{1}{c}{\textbf{$\mathcal{G}_s$}} & \multicolumn{1}{c}{\textbf{$\mathcal{G}_d$}} &  &      & \multicolumn{1}{c}{\textbf{$\mathcal{G}_s$}} & \multicolumn{1}{c}{\textbf{$\mathcal{G}_d$}} \\ \hline
\multirow{3}{*}{Birthday}  & -       &  705377  &  1028398  & \multirow{3}{*}{Fabien} & -  &  87284  &  259209  \\

 & 1      &  62457 \scriptsize{\textcolor{red}{(-91.1\%)}}  &  53578 \scriptsize{\textcolor{red}{(-94.8\%)}}  &  & 1 & 7010 \scriptsize{\textcolor{red}{(-92.0\%)}}    &   2455 \scriptsize{\textcolor{red}{(-99.1\%)}}  \\
 
 & 6      &  419283 \scriptsize{\textcolor{red}{(-40.6\%)}}   &  344054 \scriptsize{\textcolor{red}{(-66.5\%)}}  &  & 6 & 47383 \scriptsize{\textcolor{red}{(-45.7\%)}}   &  60256 \scriptsize{\textcolor{red}{(-76.8\%)}}  \\ \hline

\multicolumn{1}{l}{\multirow{3}{*}{Train}} &   -   &  1810464  &  241302  & \multirow{3}{*}{Painter}       &  -   &  103811  &  463955  \\

\multicolumn{1}{l}{}       &    1       &   57078 \scriptsize{\textcolor{red}{(-96.8\%)}}    &   10013 \scriptsize{\textcolor{red}{(-95.9\%)}}  &   &   1   & 4342 \scriptsize{\textcolor{red}{(-95.8\%)}}   &   6307 \scriptsize{\textcolor{red}{(-98.6\%)}} \\

\multicolumn{1}{l}{}       &     6      &  1221140 \scriptsize{\textcolor{red}{(-32.6\%)}}   &  57536 \scriptsize{\textcolor{red}{(-76.2\%)}}  &  &   6   & 58010 \scriptsize{\textcolor{red}{(-44.1\%)}}   & 103793\scriptsize{\textcolor{red}{(-77.6\%)}}  \\

\hline
\end{tabular}
\label{tab:ratio}
\end{table}

}

\newcommand{\tabLightFourGS}{
\begin{table}[h!]
\centering
\caption{\textbf{Direct comparison with Light4GS on the N3V dataset, following their experimental protocol.} Our method excels in high-compression and rendering speed, while Light4GS achieves higher peak PSNR.}
\label{tab:appendix_light4gs}
\begin{tabular}{lccc}
\toprule
\textbf{Model} & \textbf{PSNR (dB)} & \textbf{Size (MB)} & \textbf{FPS} \\
\midrule
Ours (High-Compression) & 27.90 & 1.48 & $\sim$160 \\
Ours (High-Quality)     & 30.80 & 12.56 & $\sim$100 \\
\midrule
Light4GS (High-Compression) & 31.48 & 3.77 & $\sim$40 \\
Light4GS (High-Quality)     & 31.69 & 5.46 & $\sim$37 \\
\bottomrule
\end{tabular}
\end{table}
}

\newcommand{\tabAblation}{
\begin{wraptable}{r}{0.5\linewidth}
\caption{\textbf{Ablation on quantizing opacity components ($\alpha^s$, $\alpha^d$, $a_*^o$, $b_*^o$) for compression levels 1 and 6 on the N3V dataset.} PSNR (dB) and relative model size reduction (\%) are reported. $\checkmark$ indicates the component is quantized.}
\centering
\scriptsize
\begin{tabular}{cccccc}
\toprule
$\alpha^s$ & $\alpha^d$ & $a_*^o$ & $b_*^o$ & $ \level{L}1$ & $\level{L}6$ \\ \hline
-& -& -& -& 26.86 (0.00\%) & 29.57 (0.00\%) \\
$\checkmark$ & & & & 26.86 (4.16\%) & 29.57 (4.01\%)\\
$\checkmark$ & $\checkmark$ & & & 26.86 (3.84\%) & 29.57 (3.39\%)\\
$\checkmark$ & $\checkmark$ & $\checkmark$ & & 26.86 (4.97\%) & 29.57 (5.30\%)\\
$\checkmark$ & $\checkmark$ &  & $\checkmark$& 26.17 (5.16\%) & 28.52 (5.58\%)\\
$\checkmark$ & $\checkmark$ & $\checkmark$ & $\checkmark$ & 26.17 (5.98\%) & 28.52 (6.87\%)\\

\hline
\end{tabular}
\label{tab:ablation}
\end{wraptable}
}

\newcommand{\tabAllAppendix}{

\begin{table}[h!]
\centering
\caption{\textbf{Quantitative results averaged over the N3V and Technicolor datasets.} We compare the proposed method (L1-L6) against the Ex4DGS~\cite{ex4dgs} using standard quality metrics and model size.}
\scriptsize
\begin{tabular}{ccccc ccccc} 
\toprule
\multicolumn{5}{c}{[N3V] Coffee Martini} & \multicolumn{5}{c}{[N3V] Cook Spinach} \\ \hline
& SSIM & PSNR & LPIPS & Size(MB) & & SSIM & PSNR & LPIPS & Size(MB) \\ \midrule
Ex4DGS~\cite{ex4dgs} & 0.918 & 28.79 & 0.070 & 129.77 & Ex4DGS~\cite{ex4dgs} & 0.956 & 33.24 & 0.042 & 118.38 \\
Ours [$\level{L}$6] & 0.897 & 26.17 & 0.091 & 15.29 \scriptsize{\textcolor{red}{(-88.2\%)}} & Ours [$\level{L}$6] & 0.942 & 30.69 & 0.064 & 10.44 \scriptsize{\textcolor{red}{(-91.2\%)}} \\
Ours [$\level{L}$5] & 0.894 & 25.92 & 0.098 & 10.60 \scriptsize{\textcolor{red}{(-91.8\%)}} & Ours [$\level{L}$5] & 0.939 & 30.50 & 0.069 & 6.75 \scriptsize{\textcolor{red}{(-94.3\%)}} \\
Ours [$\level{L}$4] & 0.892 & 25.82 & 0.101 & 7.40 \scriptsize{\textcolor{red}{(-94.3\%)}} & Ours [$\level{L}$4] & 0.935 & 30.11 & 0.075 & 4.50 \scriptsize{\textcolor{red}{(-96.2\%)}} \\
Ours [$\level{L}$3] & 0.888 & 25.74 & 0.108 & 5.21 \scriptsize{\textcolor{red}{(-96.0\%)}} & Ours [$\level{L}$3] & 0.930 & 29.55 & 0.083 & 3.12 \scriptsize{\textcolor{red}{(-97.4\%)}} \\
Ours [$\level{L}$2] & 0.882 & 25.48 & 0.119 & 3.48 \scriptsize{\textcolor{red}{(-97.3\%)}} & Ours [$\level{L}$2] & 0.922 & 28.99 & 0.099 & 2.06 \scriptsize{\textcolor{red}{(-98.3\%)}} \\
Ours [$\level{L}$1] & 0.868 & 24.97 & 0.144 & 1.92 \scriptsize{\textcolor{red}{(-98.5\%)}} & Ours [$\level{L}$1] & 0.910 & 28.38 & 0.121 & 1.15 \scriptsize{\textcolor{red}{(-99.0\%)}} \\ \hline \hline

\multicolumn{5}{c}{[N3V] Cut Roasted Beef} & \multicolumn{5}{c}{[N3V] Flame Salmon} \\ \hline
& SSIM & PSNR & LPIPS & Size(MB) & & SSIM & PSNR & LPIPS & Size(MB) \\ \midrule
Ex4DGS~\cite{ex4dgs} & 0.958 & 33.73 & 0.040 & 122.63 & Ex4DGS~\cite{ex4dgs} & 0.926 & 29.29 & 0.066 & 128.11 \\
Ours [$\level{L}$6] & 0.943 & 31.45 & 0.064 & 10.04 \scriptsize{\textcolor{red}{(-91.8\%)}} & Ours [$\level{L}$6] & 0.910 & 26.58 & 0.079 & 13.43 \scriptsize{\textcolor{red}{(-89.5\%)}} \\
Ours [$\level{L}$5] & 0.940 & 30.86 & 0.068 & 6.59 \scriptsize{\textcolor{red}{(-94.6\%)}} & Ours [$\level{L}$5] & 0.908 & 26.35 & 0.082 & 8.97 \scriptsize{\textcolor{red}{(-93.0\%)}} \\
Ours [$\level{L}$4] & 0.936 & 30.23 & 0.074 & 4.42 \scriptsize{\textcolor{red}{(-96.4\%)}} & Ours [$\level{L}$4] & 0.905 & 25.92 & 0.088 & 6.07 \scriptsize{\textcolor{red}{(-95.3\%)}} \\
Ours [$\level{L}$3] & 0.931 & 29.82 & 0.082 & 3.03 \scriptsize{\textcolor{red}{(-97.5\%)}} & Ours [$\level{L}$3] & 0.902 & 25.68 & 0.095 & 4.18 \scriptsize{\textcolor{red}{(-96.7\%)}} \\
Ours [$\level{L}$2] & 0.924 & 29.04 & 0.093 & 2.01 \scriptsize{\textcolor{red}{(-98.4\%)}} & Ours [$\level{L}$2] & 0.893 & 24.87 & 0.107 & 2.77 \scriptsize{\textcolor{red}{(-97.8\%)}} \\
Ours [$\level{L}$1] & 0.911 & 28.02 & 0.122 & 1.12 \scriptsize{\textcolor{red}{(-99.1\%)}} & Ours [$\level{L}$1] & 0.875 & 24.16 & 0.138 & 1.52 \scriptsize{\textcolor{red}{(-98.8\%)}} \\ \hline \hline

\multicolumn{5}{c}{[N3V] Flame Steak} & \multicolumn{5}{c}{[N3V] Sear Steak} \\ \hline
& SSIM & PSNR & LPIPS & Size(MB) & & SSIM & PSNR & LPIPS & Size(MB) \\ \midrule
Ex4DGS~\cite{ex4dgs} & 0.963 & 33.91 & 0.034 & 101.18 & Ex4DGS~\cite{ex4dgs} & 0.960 & 33.69 & 0.035 & 93.69 \\
Ours [$\level{L}$6] & 0.950 & 31.34 & 0.059 & 8.49 \scriptsize{\textcolor{red}{(-91.6\%)}} & Ours [$\level{L}$6] & 0.950 & 31.70 & 0.056 & 8.66 \scriptsize{\textcolor{red}{(-90.8\%)}} \\
Ours [$\level{L}$5] & 0.947 & 30.86 & 0.063 & 5.41 \scriptsize{\textcolor{red}{(-94.7\%)}} & Ours [$\level{L}$5] & 0.948 & 31.22 & 0.061 & 5.43 \scriptsize{\textcolor{red}{(-94.2\%)}} \\
Ours [$\level{L}$4] & 0.943 & 30.38 & 0.070 & 3.56 \scriptsize{\textcolor{red}{(-96.5\%)}} & Ours [$\level{L}$4] & 0.944 & 30.62 & 0.068 & 3.58 \scriptsize{\textcolor{red}{(-96.2\%)}} \\
Ours [$\level{L}$3] & 0.938 & 29.95 & 0.080 & 2.40 \scriptsize{\textcolor{red}{(-97.6\%)}} & Ours [$\level{L}$3] & 0.938 & 29.95 & 0.077 & 2.48 \scriptsize{\textcolor{red}{(-97.3\%)}} \\
Ours [$\level{L}$2] & 0.931 & 29.13 & 0.093 & 1.58 \scriptsize{\textcolor{red}{(-98.4\%)}} & Ours [$\level{L}$2] & 0.931 & 29.34 & 0.093 & 1.64 \scriptsize{\textcolor{red}{(-98.3\%)}} \\
Ours [$\level{L}$1] & 0.918 & 28.33 & 0.122 & 0.89 \scriptsize{\textcolor{red}{(-99.1\%)}} & Ours [$\level{L}$1] & 0.917 & 28.39 & 0.122 & 0.94 \scriptsize{\textcolor{red}{(-99.0\%)}} \\ \hline 
\hline

\multicolumn{5}{c}{} & \multicolumn{5}{c}{[Technicolor] Train} \\ \cline{6-10}
 & & & & &  & SSIM & PSNR & LPIPS & Size(MB)\\ \cline{6-10}
 & & & & & Ex4DGS~\cite{ex4dgs} & 0.928 & 31.37 & 0.055 & 216.17 \\
& & & & & Ours [$\level{L}$6] & 0.931 & 31.55 & 0.040 & 33.19 \scriptsize{\textcolor{red}{(-84.6\%)}} \\
& & & & &  Ours [$\level{L}$5] & 0.928 & 31.38 & 0.045 & 21.76 \scriptsize{\textcolor{red}{(-89.9\%)}} \\
& & & & &  Ours [$\level{L}$4] & 0.913 & 30.39 & 0.060 & 12.73 \scriptsize{\textcolor{red}{(-94.1\%)}} \\
& & & & &  Ours [$\level{L}$3] & 0.889 & 29.86 & 0.092 & 7.61 \scriptsize{\textcolor{red}{(-96.5\%)}} \\
& & & & &  Ours [$\level{L}$2] & 0.845 & 28.44 & 0.146 & 4.55 \scriptsize{\textcolor{red}{(-97.9\%)}} \\
& & & & & Ours [$\level{L}$1] & 0.765 & 26.32 & 0.247 & 2.32 \scriptsize{\textcolor{red}{(-98.9\%)}} \\ \hline

\multicolumn{5}{c}{[Technicolor] Birthday} & \multicolumn{5}{c}{[Technicolor] Fabien} \\ \hline
& SSIM & PSNR & LPIPS & Size(MB) & & SSIM & PSNR & LPIPS & Size(MB) \\ \midrule
Ex4DGS~\cite{ex4dgs} & 0.942 & 32.38 & 0.044 & 162.38 & Ex4DGS~\cite{ex4dgs} & 0.897 & 35.38 & 0.123 & 83.20 \\
Ours [$\level{L}$6] & 0.932 & 31.33 & 0.046 & 33.62 \scriptsize{\textcolor{red}{(-79.3\%)}} & Ours [$\level{L}$6] & 0.871 & 32.90 & 0.168 & 5.89 \scriptsize{\textcolor{red}{(-92.9\%)}} \\
Ours [$\level{L}$5] & 0.932 & 31.33 & 0.047 & 25.47 \scriptsize{\textcolor{red}{(-84.3\%)}} & Ours [$\level{L}$5] & 0.867 & 32.74 & 0.182 & 2.86 \scriptsize{\textcolor{red}{(-96.6\%)}} \\
Ours [$\level{L}$4] & 0.929 & 31.08 & 0.051 & 18.50 \scriptsize{\textcolor{red}{(-88.6\%)}} & Ours [$\level{L}$4] & 0.861 & 32.60 & 0.200 & 1.77 \scriptsize{\textcolor{red}{(-97.9\%)}} \\
Ours [$\level{L}$3] & 0.924 & 31.00 & 0.057 & 13.59 \scriptsize{\textcolor{red}{(-91.6\%)}} & Ours [$\level{L}$3] & 0.856 & 32.06 & 0.214 & 1.27 \scriptsize{\textcolor{red}{(-98.5\%)}} \\
Ours [$\level{L}$2] & 0.916 & 30.56 & 0.068 & 9.34 \scriptsize{\textcolor{red}{(-94.2\%)}} & Ours [$\level{L}$2] & 0.850 & 31.68 & 0.228 & 0.90 \scriptsize{\textcolor{red}{(-98.9\%)}} \\
Ours [$\level{L}$1] & 0.896 & 29.46 & 0.092 & 5.39 \scriptsize{\textcolor{red}{(-96.7\%)}} & Ours [$\level{L}$1] & 0.840 & 30.44 & 0.254 & 0.58 \scriptsize{\textcolor{red}{(-99.3\%)}} \\ \hline \hline

\multicolumn{5}{c}{[Technicolor] Painter} & \multicolumn{5}{c}{[Technicolor] Theater} \\ \hline
& SSIM & PSNR & LPIPS & Size(MB) & & SSIM & PSNR & LPIPS & Size(MB) \\ \midrule
Ex4DGS~\cite{ex4dgs} & 0.932 & 36.73 & 0.091 & 106.30 & Ex4DGS~\cite{ex4dgs} & 0.883 & 31.84 & 0.129 & 132.89 \\
Ours [$\level{L}$6] & 0.914 & 35.05 & 0.110 & 9.19 \scriptsize{\textcolor{red}{(-91.4\%)}} & Ours [$\level{L}$6] & 0.871 & 30.16 & 0.138 & 16.27 \scriptsize{\textcolor{red}{(-87.8\%)}} \\
Ours [$\level{L}$5] & 0.905 & 34.00 & 0.122 & 5.55 \scriptsize{\textcolor{red}{(-94.8\%)}} & Ours [$\level{L}$5] & 0.869 & 30.12 & 0.150 & 9.31 \scriptsize{\textcolor{red}{(-93.0\%)}} \\
Ours [$\level{L}$4] & 0.891 & 33.29 & 0.142 & 3.47 \scriptsize{\textcolor{red}{(-96.7\%)}} & Ours [$\level{L}$4] & 0.858 & 29.70 & 0.172 & 5.58 \scriptsize{\textcolor{red}{(-95.8\%)}} \\
Ours [$\level{L}$3] & 0.873 & 32.20 & 0.168 & 2.30 \scriptsize{\textcolor{red}{(-97.8\%)}} & Ours [$\level{L}$3] & 0.849 & 29.42 & 0.191 & 3.57 \scriptsize{\textcolor{red}{(-97.3\%)}} \\
Ours [$\level{L}$2] & 0.848 & 30.96 & 0.210 & 1.47 \scriptsize{\textcolor{red}{(-98.6\%)}} & Ours [$\level{L}$2] & 0.833 & 29.00 & 0.221 & 2.19 \scriptsize{\textcolor{red}{(-98.4\%)}} \\
Ours [$\level{L}$1] & 0.804 & 28.93 & 0.293 & 0.79 \scriptsize{\textcolor{red}{(-99.3\%)}} & Ours [$\level{L}$1] & 0.804 & 27.84 & 0.274 & 1.16 \scriptsize{\textcolor{red}{(-99.1\%)}} \\ 

\bottomrule
\end{tabular}
\label{appendix:all_tab}
\end{table}
}

\newcommand{\tabStorageApp}{

\begin{table}[h!]
\centering
\caption{\textbf{Storage allocation (MB and percentage) for components in our RD-optimized 4DGS model.} Technicolor Scenes at Low ($\level{L}$1) and High ($\level{L}$6) rate settings.}
\scriptsize
\resizebox{\linewidth}{!}{

\begin{tabular}{ccccc ccccc}
\hline
\multicolumn{5}{c}{\textbf{Train}} & \multicolumn{5}{c}{\textbf{Theater}}\\ \hline
& \multicolumn{2}{c}{Low Rate [$\level{L}$1]}  & \multicolumn{2}{c}{High Rate [$\level{L}$6] } & & \multicolumn{2}{c}{Low Rate [$\level{L}$1]}  & \multicolumn{2}{c}{High Rate [$\level{L}$6] } \\
\hline
Component & Rate (MB) & Ratio (\%) & Rate (MB) & Ratio (\%) & Component & Rate (MB) & Ratio (\%) & Rate (MB) & Ratio (\%) \\ \hline

sort idx & 0.537 & 22.08\% & 10.230 & 29.40\% & $\mathbf{F}_{\text{masked}}$ & 0.242 & 19.86\% & 5.660 & 33.17\% \\
Indexes & 0.458 & 18.81\% & 7.327 & 21.06\% & Codebooks & 0.231 & 18.96\% & 4.272 & 25.04\% \\
$\mathbf{F}_{\text{masked}}$ & 0.361 & 14.84\% & 7.327 & 21.06\% & Indexes & 0.222 & 18.24\% & 2.698 & 15.81\% \\
$\boldsymbol{\mu}_{disp}$ & 0.343 & 14.09\% & 6.394 & 18.37\% & sort idx & 0.189 & 15.51\% & 1.258 & 7.37\% \\
$\boldsymbol{\mu}_0$ & 0.343 & 14.09\% & 2.072 & 5.95\% & $\boldsymbol{\mu}_{disp}$ & 0.101 & 8.33\% & 1.080 & 6.33\% \\
Codebooks & 0.230 & 9.46\% & 0.866 & 2.49\% & $\boldsymbol{\mu}_0$ & 0.101 & 8.33\% & 1.080 & 6.33\% \\
Logits & 0.080 & 3.31\% & 0.461 & 1.32\% & Logits & 0.077 & 6.33\% & 0.883 & 5.18\% \\
$\beta_*^o$ & 0.080 & 3.31\% & 0.123 & 0.35\% & $\beta_*^o$ & 0.054 & 4.43\% & 0.130 & 0.76\% \\
Header & 0.000 & 0.00\% & 0.000 & 0.00\% & Header & 0.000 & 0.00\% & 0.000 & 0.00\% \\ \hline
Total & 2.319 & - & 33.187 & - & Total & 1.162 & - & 16.271 & - \\
\hline

\hline 
\hline
\multicolumn{5}{c}{\textbf{Painter}} & \multicolumn{5}{c}{\textbf{Fabien}}\\ \hline

& \multicolumn{2}{c}{Low Rate [$\level{L}$1]}  & \multicolumn{2}{c}{High Rate [$\level{L}$6] } & & \multicolumn{2}{c}{Low Rate [$\level{L}$1]}  & \multicolumn{2}{c}{High Rate [$\level{L}$6] } \\
\hline
Component & Rate (MB) & Ratio (\%) & Rate (MB) & Ratio (\%) & Component & Rate (MB) & Ratio (\%) & Rate (MB) & Ratio (\%) \\ \hline

$\mathbf{F}_{\text{masked}}$ & 0.227 & 27.48\% & 3.737 & 38.79\% &Codebooks & 0.188 & 31.09\% & 2.169 & 35.11\% \\
Codebooks & 0.196 & 23.66\% & 2.499 & 25.94\% &$\mathbf{F}_{\text{masked}}$ & 0.089 & 14.67\% & 1.516 & 24.53\% \\
Indexes & 0.150 & 18.12\% & 1.295 & 13.44\% &Indexes & 0.084 & 13.89\% & 0.861 & 13.94\% \\
sort idx & 0.085 & 10.33\% & 0.831 & 8.62\% &sort idx & 0.076 & 12.57\% & 0.482 & 7.81\% \\
Logits & 0.065 & 7.91\% & 0.472 & 4.90\% &Logits & 0.063 & 10.48\% & 0.480 & 7.77\% \\
$\beta_*^o$ & 0.051 & 6.13\% & 0.348 & 3.62\% &$\boldsymbol{\mu}_{disp}$ & 0.042 & 7.00\% & 0.285 & 4.61\% \\
$\boldsymbol{\mu}_{disp}$ & 0.026 & 3.18\% & 0.348 & 3.62\% &$\boldsymbol{\mu}_0$ & 0.042 & 7.00\% & 0.285 & 4.61\% \\
$\boldsymbol{\mu}_0$ & 0.026 & 3.18\% & 0.104 & 1.08\% &$\beta_*^o$ & 0.020 & 3.29\% & 0.100 & 1.62\% \\
Header & 0.000 & 0.01\% & 0.000 & 0.00\% &Header & 0.000 & 0.01\% & 0.000 & 0.00\% \\ \hline
Total & 0.789 & - & 9.188 & - & Total & 0.576 & - & 5.892 & - \\
\hline 

\hline
\multicolumn{5}{c}{\textbf{Birthday}} & \multicolumn{5}{c}{}\\ \cline{1-5}
& \multicolumn{2}{c}{Low Rate [$\level{L}$1]}  & \multicolumn{2}{c}{High Rate [$\level{L}$6] } \\
\cline{1-5}
Component & Rate (MB) & Ratio (\%) & Rate (MB) & Ratio (\%) \\ \cline{1-5}
$\mathbf{F}_{\text{masked}}$ & 1.929 & 34.13\% & 12.386 & 35.14\% \\
Indexes & 1.280 & 22.65\% & 8.057 & 22.86\% \\
sort idx & 0.929 & 16.43\% & 6.107 & 17.32\% \\
$\beta_*^o$ & 0.429 & 7.59\% & 2.753 & 7.81\% \\
$\boldsymbol{\mu}_{disp}$ & 0.375 & 6.63\% & 2.516 & 7.14\% \\
$\boldsymbol{\mu}_0$ & 0.375 & 6.63\% & 2.516 & 7.14\% \\
Codebooks & 0.250 & 4.43\% & 0.794 & 2.25\% \\
Logits & 0.085 & 1.50\% & 0.122 & 0.35\% \\
Header & 0.000 & 0.00\% & 0.000 & 0.00\% \\ \cline{1-5}
Total & 5.391 & - & 33.617 & - \\
\hline \\
\end{tabular}
}

\label{appendix:storagetechni}
\end{table}

}

\newcommand{\tabStorageNtVApp}{

\begin{table}[h!]
\small
\centering
\caption{\textbf{Storage allocation (MB and percentage) for components in our RD-optimized 4DGS model.} N3V Scenes at Low ($\level{L}$1) and High ($\level{L}$6) rate settings.}
\scriptsize
\resizebox{\linewidth}{!}{
\begin{tabular}{ccccc ccccc}

\hline
\multicolumn{5}{c}{\textbf{Coffee Martini}} & \multicolumn{5}{c}{\textbf{Cook Spinach}} \\ \hline

& \multicolumn{2}{c}{Low Rate [$\level{L}$1]}  & \multicolumn{2}{c}{High Rate [$\level{L}$6] }  & & \multicolumn{2}{c}{Low Rate [$\level{L}$1]}  & \multicolumn{2}{c}{High Rate [$\level{L}$6] } \\
\hline
Component & Rate (MB) & Proportion (\%) & Rate (MB) & Proportion (\%) &Component & Rate (MB) & Proportion (\%) & Rate (MB) & Proportion (\%) \\ \hline
$\mathbf{F}_{\text{masked}}$ & 0.448 & 22.21\% & 4.574 & 28.53\% &$\mathbf{F}_{\text{masked}}$ & 0.285 & 23.71\% & 3.699 & 33.78\% \\
Indexes & 0.413 & 20.46\% & 4.051 & 25.26\% &Codebooks & 0.241 & 20.01\% & 2.869 & 26.20\% \\
sort idx & 0.319 & 15.81\% & 2.574 & 16.05\% &Indexes & 0.234 & 19.41\% & 1.409 & 12.87\% \\
Codebooks & 0.284 & 14.08\% & 1.676 & 10.45\% &sort idx & 0.151 & 12.50\% & 0.862 & 7.87\% \\
$\boldsymbol{\mu}_0$ & 0.214 & 10.62\% & 1.676 & 10.45\% &$\boldsymbol{\mu}_0$ & 0.097 & 8.07\% & 0.851 & 7.77\% \\
$\boldsymbol{\mu}_{disp}$ & 0.214 & 10.62\% & 0.999 & 6.23\% &$\boldsymbol{\mu}_{disp}$ & 0.097 & 8.07\% & 0.851 & 7.77\% \\
Logits & 0.091 & 4.53\% & 0.339 & 2.11\% &Logits & 0.078 & 6.46\% & 0.274 & 2.50\% \\
$\beta_*^o$ & 0.033 & 1.66\% & 0.146 & 0.91\% &$\beta_*^o$ & 0.021 & 1.78\% & 0.135 & 1.24\% \\
Header & 0.000 & 0.00\% & 0.000 & 0.00\% &Header & 0.000 & 0.00\% & 0.000 & 0.00\% \\ \hline
Total & 1.923 & - & 15.292 & - & Total & 1.149 & - & 10.444 & - \\ \hline

\hline
\multicolumn{5}{c}{\textbf{Cut Roasted Beef}} & \multicolumn{5}{c}{\textbf{Flame Salmon}}\\ \hline

& \multicolumn{2}{c}{Low Rate [$\level{L}$1]}  & \multicolumn{2}{c}{High Rate [$\level{L}$6] } & & \multicolumn{2}{c}{Low Rate [$\level{L}$1]}  & \multicolumn{2}{c}{High Rate [$\level{L}$6] } \\
\hline
Component & Rate (MB) & Proportion (\%) & Rate (MB) & Proportion (\%) & Component & Rate (MB) & Proportion (\%) & Rate (MB) & Proportion (\%) \\ \hline

$\mathbf{F}_{\text{masked}}$ & 0.262 & 22.34\% & 3.396 & 32.25\% &$\mathbf{F}_{\text{masked}}$ & 0.303 & 19.03\% & 3.809 & 27.05\% \\
Codebooks & 0.238 & 20.29\% & 2.699 & 25.62\% &Indexes & 0.303 & 18.98\% & 3.453 & 24.51\% \\
Indexes & 0.222 & 18.88\% & 1.432 & 13.60\% &Codebooks & 0.271 & 17.01\% & 2.339 & 16.61\% \\
sort idx & 0.154 & 13.08\% & 0.885 & 8.41\% &sort idx & 0.256 & 16.06\% & 1.542 & 10.95\% \\
$\boldsymbol{\mu}_0$ & 0.101 & 8.58\% & 0.885 & 8.41\% &$\boldsymbol{\mu}_0$ & 0.175 & 10.99\% & 1.542 & 10.95\% \\
$\boldsymbol{\mu}_{disp}$ & 0.101 & 8.58\% & 0.849 & 8.06\% &$\boldsymbol{\mu}_{disp}$ & 0.175 & 10.99\% & 0.973 & 6.91\% \\
Logits & 0.077 & 6.57\% & 0.252 & 2.39\% &Logits & 0.088 & 5.51\% & 0.282 & 2.01\% \\
$\beta_*^o$ & 0.020 & 1.68\% & 0.133 & 1.27\% &$\beta_*^o$ & 0.023 & 1.43\% & 0.143 & 1.02\% \\
Header & 0.000 & 0.00\% & 0.000 & 0.00\% &Header & 0.000 & 0.00\% & 0.000 & 0.00\% \\ \hline
Total & 1.120 & - & 10.044 & - & Total & 1.521 & - & 13.432 & - \\ \hline

\hline
\multicolumn{5}{c}{\textbf{Flame Steak}} & \multicolumn{5}{c}{\textbf{Sear Steak}}\\ \hline
& \multicolumn{2}{c}{Low Rate [$\level{L}$1]}  & \multicolumn{2}{c}{High Rate [$\level{L}$6] } & & \multicolumn{2}{c}{Low Rate [$\level{L}$1]}  & \multicolumn{2}{c}{High Rate [$\level{L}$6] } \\
\hline
Component & Rate (MB) & Proportion (\%) & Rate (MB) & Proportion (\%) & Component & Rate (MB) & Proportion (\%) & Rate (MB) & Proportion (\%) \\ \hline
Codebooks & 0.222 & 23.68\% & 2.680 & 30.10\% &Codebooks & 0.225 & 22.82\% & 2.738 & 30.15\% \\
$\mathbf{F}_{\text{masked}}$ & 0.183 & 19.54\% & 2.200 & 24.71\% &$\mathbf{F}_{\text{masked}}$ & 0.189 & 19.21\% & 2.278 & 25.08\% \\
Indexes & 0.164 & 17.49\% & 1.263 & 14.18\% &Indexes & 0.171 & 17.31\% & 1.282 & 14.12\% \\
sort idx & 0.121 & 12.90\% & 0.834 & 9.36\% &sort idx & 0.134 & 13.56\% & 0.830 & 9.14\% \\
$\boldsymbol{\mu}_0$ & 0.090 & 9.11\% & 0.810 & 8.91\% &$\boldsymbol{\mu}_0$ & 0.081 & 8.60\% & 0.798 & 8.97\% \\
$\boldsymbol{\mu}_{disp}$ & 0.081 & 8.60\% & 0.798 & 8.97\% &$\boldsymbol{\mu}_{disp}$ & 0.090 & 9.11\% & 0.810 & 8.91\% \\
Logits & 0.072 & 7.70\% & 0.199 & 2.23\% &Logits & 0.073 & 7.43\% & 0.203 & 2.24\% \\
$\beta_*^o$ & 0.014 & 1.47\% & 0.132 & 1.48\% &$\beta_*^o$ & 0.014 & 1.45\% & 0.131 & 1.44\% \\
Header & 0.000 & 0.01\% & 0.000 & 0.00\% &Header & 0.000 & 0.01\% & 0.000 & 0.00\% \\ \hline
Total & 0.895 & - & 8.491 & - &Total & 0.941 & - & 8.661 & - \\ \hline
\end{tabular}
}
\label{appendix:storagen3v}
\end{table}

}

\newcommand{\tabMultiLevelWavelet}{
\begin{wraptable}{r}{0.55\linewidth}
\centering
\vspace{-4mm}
\caption{\textbf{Ablation study on multi-level wavelet decomposition for dynamic positions, averaged over the N3V dataset.} Columns compare decompositions retaining different fractions of wavelet coefficients.}
\vspace{1mm}
\scriptsize
\begin{tabular}{lccc}
\toprule
\textbf{Compression} & \textbf{Level 1. (1/2)} & \textbf{Level 2. (1/4)} & \textbf{Level 3. (1/8)} \\
\textbf{Level} & \textbf{PSNR / Size} & \textbf{PSNR / Size} & \textbf{PSNR / Size} \\
\midrule
$\level{L}$6 & 30.26 / 14.67 & 28.47 / 11.35 & 27.74 / 9.80 \\
$\level{L}$5 & 29.59 / 9.14  & 28.57 / 7.51  & 27.72 / 6.50 \\
$\level{L}$4 & 29.14 / 6.23  & 28.14 / 5.14  & 27.44 / 4.44 \\
$\level{L}$3 & 28.59 / 4.37  & 27.81 / 3.61  & 27.38 / 3.10 \\
$\level{L}$2 & 28.27 / 2.91  & 27.57 / 2.42  & 27.17 / 2.05 \\
$\level{L}$1 & 27.39 / 1.63  & 26.89 / 1.36  & 26.43 / 1.16 \\
\bottomrule
\end{tabular}
\label{tab:ablation_wavelet_multilevel}
\end{wraptable}
}

\newcommand{\tabHigherFid}{
\begin{wraptable}{r}{0.55\linewidth}
\centering
\vspace{-4mm}
\caption{\textbf{Analysis of the RD trade-off at higher fidelities.} We compare L6 configuration against variants with a reduced ($\lambda_\text{R}$) and an increased codebook size. Results are averaged over the N3V dataset.\protect\footnotemark}
\vspace{1mm}
\label{tab:ablation_high_fidelity}
\scriptsize
\begin{tabular}{lcc}
\toprule
\textbf{Configuration} & \textbf{Avg PSNR (dB)} & \textbf{Size (MB)} \\
\midrule
Ex4DGS (Baseline) & 32.11 & 115.63 \\
\midrule
Ours ($\level{L}$6, low $\lambda_\text{R}$, large codebook) & 30.29 & 27.21 \\
Ours ($\level{L}$6, low $\lambda_\text{R}$) & 30.04 & 14.40 \\
\midrule
Ours ($\level{L}$6) & 30.26 & 14.67 \\
Ours ($\level{L}$1) & 27.39 & 1.63 \\
\bottomrule
\end{tabular}
\end{wraptable}
}
\usepackage[capitalize]{cleveref}
\crefname{section}{Sec.}{Secs.}
\Crefname{section}{Section}{Sections}
\Crefname{table}{Table}{Tables}
\crefname{table}{Tab.}{Tabs.}

\newcommand{\level}[1]{\mathbb{#1}}

\title{Temporal Smoothness-Aware Rate-Distortion Optimized 4D Gaussian Splatting}

\author{%
  Hyeongmin Lee \\
  Twelve Labs \\
  \texttt{lucas@twelvelabs.io} \\
  \And
  Kyungjune Baek\thanks{Corresponding Author} \\
  Sejong University \\
  \texttt{kyungjune.baek@sejong.ac.kr} \\
}

\begin{document}

\maketitle

\begin{abstract}
Dynamic 4D Gaussian Splatting (4DGS) effectively extends the high-speed rendering capabilities of 3D Gaussian Splatting (3DGS) to represent volumetric videos. However, the large number of Gaussians, substantial temporal redundancies, and especially the absence of an entropy-aware compression framework result in large storage requirements. Consequently, this poses significant challenges for practical deployment, efficient edge-device processing, and data transmission.
In this paper, we introduce a novel end-to-end RD-optimized compression framework tailored for 4DGS, aiming to enable flexible, high-fidelity rendering across varied computational platforms.
Leveraging Fully Explicit Dynamic Gaussian Splatting (Ex4DGS), one of the state-of-the-art 4DGS methods, as our baseline, we start from the existing 3DGS compression methods for compatibility while effectively addressing additional challenges introduced by the temporal axis. In particular, instead of storing motion trajectories independently per point, we employ a wavelet transform to reflect the real-world smoothness prior, significantly enhancing storage efficiency.
This approach yields significantly improved compression ratios and provides a user-controlled balance between compression efficiency and rendering quality. Extensive experiments demonstrate the effectiveness of our method, achieving up to 91$\times$ compression compared to the original Ex4DGS model while maintaining high visual fidelity. These results highlight the applicability of our framework for real-time dynamic scene rendering in diverse scenarios, from resource-constrained edge devices to high-performance environments. The source code is available at \url{https://github.com/HyeongminLEE/RD4DGS}.

\end{abstract}

\section{Introduction}
\label{sec:intro}
The advent of 3D Gaussian Splatting (3DGS)~\cite{3dgs} marked a significant step in real-time radiance field rendering, offering superior speed and simpler training compared to Neural Radiance Field (NeRF)-based approaches~\cite{nerf,nerfies,wu2023efficient,dihlmann2024signerf,zhong2024structured,ren2024nerf}. Furthermore, the Gaussian Splatting framework is not restricted to static scenes but has been effectively extended to represent dynamic, or 4D, scenes such as videos by capturing temporal variations~\cite{4dgs,yang2023real,yang2024deformable,duisterhof2023deformgs,duisterhof2023md, girish2024queen}.

Dynamic 4D Gaussian Splatting (4DGS) methods typically require a substantial number of Gaussians and extensive Spherical Harmonics (SH) parameters to represent volumetric video accurately. Additionally, dynamic points in these methods are often stored independently per timestamp, inherently causing significant temporal redundancies. Consequently, such characteristics lead to large models and excessive storage demands, severely restricting applicability in resource-constrained scenarios, including edge computing and efficient data transmission. Although several existing techniques~\cite{liu2025light4gs,yuan20251000+,zhang2024mega,liu2024lgs} have explored model size reduction, there is currently no entropy-aware, bit-level compression method for 4DGS that enables flexible rate-distortion optimization.

In this paper, we propose a novel rate-distortion (RD) optimized compression method specifically designed for 4D Gaussian Splatting (4DGS), enabling real-time applications across diverse computing environments, ranging from edge devices to general computing platforms. We adopt Fully Explicit Dynamic Gaussian Splatting (Ex4DGS)~\cite{ex4dgs} as our baseline model and introduce a novel approach that leverages inherent video characteristics beyond traditional RD-optimization. 
Specifically, following Ex4DGS, we decompose the scene into static and dynamic components. For components common to both 3DGS and 4DGS, we apply masking and vector quantization after model training, following the approach of Wang et al.~\cite{wang2024rdo3dgs}. 
Although Ex4DGS reduces redundancy by storing dynamic points only at selected keyframes instead of every frame, the positions at these keyframes are stored independently, inherently limiting the exploitation of temporal redundancy and thus reducing compression efficiency.
Instead of utilizing the compression method for 3DGS directly, we leverage the prior that each point moves along a smooth trajectory. To this end, we apply a novel wavelet transform to each point’s trajectory and discard high-frequency detail components, significantly reducing storage without severely compromising fidelity. Additionally, we conduct a fine-grained analysis, evaluating the rate-distortion impact of quantizing individual dynamic opacity elements. Based on this analysis, we selectively apply quantization only to those elements that offer a favorable trade-off between compression rate and rendering quality.

Our method achieves model sizes as low as only 1.1\% on average of the original Ex4DGS model while maintaining reasonable visual fidelity.
Furthermore, our framework allows users to flexibly control the trade-off between distortion and data rate based on specific scenario requirements.

Our contributions can be summarized as follows:

\begin{itemize}[leftmargin=3.5mm]
\item We propose a novel end-to-end rate-distortion optimized compression framework specifically designed for 4D Gaussian Splatting representations, using Ex4DGS as the baseline.
\item To the best of our knowledge, this work is the first to introduce a bit-level rate-distortion optimizable compression framework for dynamic 4D Gaussian representations.
\item We apply the pointwise wavelet transform to further compress the position trajectories of dynamic Gaussian points, effectively leveraging temporal redundancy inherent in 4D representations.
\item Our method achieves significant compression ratios (up to 91$\times$) for dynamic scenes while maintaining reasonable visual fidelity and offering flexible control over the rate-distortion trade-off suitable for diverse computing platforms.
\end{itemize}

\section{Related Work}
\label{sec:relatedwork}
\subsection{Neural Rendering}
Neural Radiance Fields (NeRF) introduced a novel view synthesis approach by learning density and color via a multi-layer perceptron from 3D coordinates and viewing directions~\cite{nerf}. Despite their effectiveness, NeRF-based methods suffer from slow rendering and training speed. Follow-up works accelerated rendering by employing multi-resolution hash grids~\cite{instantngp} or optimizing sparse voxel grids directly~\cite{fridovich2022plenoxels}, though often at the expense of increased memory usage.

3D Gaussian Splatting (3DGS)~\cite{3dgs} emerged as a fast and memory-efficient alternative by representing scenes with anisotropic 3D Gaussians rather than dense voxels or networks. By optimizing the parameters, including mean, covariance, and color, 3DGS achieves real-time rendering at high resolutions using efficient algorithms instead of volumetric ray marching. Recent studies demonstrate that 3DGS matches or surpasses NeRF in rendering quality, speed, and scalability.

Dynamic scene modeling extends this idea into 4D Gaussian Splatting (4DGS), incorporating time as an additional dimension. Dynamic NeRF methods~\cite{nerfies,hypernerf} learn deformation fields or multi-space embeddings to handle scene motion, while others utilize representations like temporal tri-planes for compression~\cite{wu2024tetrirf}, but they remain computationally heavy and slow to render. In contrast, 4DGS~\cite{4dgs,yang2024deformable} was introduced as a unified representation for spatiotemporal data, using a set of 4D Gaussians with per-Gaussian motion trajectories or a deformation network to render moving scenes in real time. Overall, Gaussian Splatting has rapidly become a promising alternative to NeRF, offering comparable quality with significantly improved rendering speed, which is crucial for interactive applications.

\subsection{Gaussian Splatting Compression}
Explicit scene representations like 3D Gaussian Splatting (3DGS)~\cite{3dgs} enable real-time, high-fidelity rendering but inherently introduce substantial memory requirements. These arise from the need to store numerous Gaussian primitives, each characterized by parameters for position, shape, opacity, and color (often represented by complex spherical harmonics). This significant memory footprint has spurred the development of various compression techniques specifically designed for 3DGS.

Recent approaches tailored for compressing 3DGS include Compact3DGS~\cite{compact3dgs}, which employs trainable masking to prune redundant Gaussians, CompGS~\cite{navaneet2024compgs}, adopting vector quantization to store Gaussians, HAC++\cite{hacpp}, leveraging hash-grid-assisted entropy coding, and Context 3DGS~\cite{wang2024contextgs}, utilizing an autoregressive, anchor-based context model for hierarchical Gaussian compression.

Extending these techniques to dynamic scenes poses greater memory challenges due to temporal variations. Recent specialized solutions include Light4GS~\cite{liu2025light4gs}, combining spatio-temporal pruning with entropy coding; 4DGS-1K~\cite{yuan20251000+}, optimizing Gaussian reuse and visibility masks for ultra-fast rendering; MEGA~\cite{zhang2024mega}, replacing spherical harmonics with lightweight predictors for significant compression; and LGS~\cite{liu2024lgs}, which prunes Gaussians and deformation fields within a framework highly specialized for surgical scenes using stereo vision. These methods highlight the growing focus on efficient and high-quality rendering of dynamic scenes. QUEEN~\cite{girish2024queen} introduces integer-based quantization enabling entropy coding; however, it does not explicitly optimize entropy within a rate-distortion framework, leaving rate-distortion optimization for dynamic Gaussian Splatting unexplored. Furthermore, QUEEN's primary objective of per-frame compression for streaming applications fundamentally differs from our goal of compressing the entire dynamic scene model for efficient storage.

\subsection{End-to-End Rate-Distortion Optimized Compression}
Rate-distortion (RD) optimization balances bitrate against reconstruction quality, a core principle extensively used in traditional and neural compression methods. End-to-end image compression frameworks by Ball\'e et al.~\cite{balle2017end} introduced autoencoder-based models jointly trained to minimize distortion and bitrate, laying the foundation for differentiable compression. Follow-up studies improved efficiency through advanced context modeling and latent priors~\cite{cheng2020learned,li2021deep,he2022elic}.

Recently, RD-optimization concepts have been applied to neural scene representations. For example, Takikawa et al. presented a neural field that continuously adjusts detail level and bitrate via a tunable parameter~\cite{takikawa2022variable}. Specifically for 3D Gaussian Splatting, Wang et al.~\cite{wang2024rdo3dgs} proposed an RD-optimization-based framework that jointly optimizes Gaussian parameters with learnable pruning and entropy-constrained quantization, achieving substantial reductions in model size with minimal quality loss. Such approaches highlight the benefit of integrating RD-optimization directly into the training process, enabling flexible compression tailored for diverse computational environments. While compression techniques exist for 4DGS and RD-optimization has been applied to 3DGS, a dedicated end-to-end RD-optimized framework specifically for 4DGS, offering flexible control over the rate-quality trade-off by effectively exploiting spatiotemporal structure, is still lacking. In this work, we develop a novel method to apply end-to-end RD-optimized compression to the 4D domain, which is related to spatiotemporal representation.

\section{Method} 
\label{sec:method}
\subsection{Preliminary}
In this section, we first introduce the fundamental concepts behind 3D Gaussian Splatting (3DGS)~\cite{3dgs}. We then briefly describe its extension to 4D scenes and detail our baseline model, Ex4DGS~\cite{ex4dgs}, which explicitly models temporal dynamics. Finally, we outline the principles of end-to-end rate-distortion optimization as applied to Gaussian Splatting.

\noindent\textbf{3D Gaussian Splatting.} 3D Gaussian Splatting (3DGS)~\cite{3dgs} represents scenes using explicit 3D Gaussian primitives. Each primitive $i$ is parameterized by a position (mean) $\boldsymbol{\mu}_i \in \mathbb{R}^3$, a covariance matrix $\boldsymbol{\Sigma}_i \in \mathbb{R}^{3\times3}$, an opacity value $\alpha_i \in [0, 1]$, and view-dependent color represented by Spherical Harmonics (SH) coefficients $\mathbf{c}_i$. The 3D Gaussian function is defined as:

\begin{equation}
G(\mathbf{x}) = e^{-\frac{1}{2}(\mathbf{x}-\boldsymbol{\mu}_i)^{T}\boldsymbol{\Sigma}_i^{-1}(\mathbf{x}-\boldsymbol{\mu}_i)} \ .
\end{equation}
The 3D covariance matrix $\boldsymbol{\Sigma}_i$ defines the shape and orientation and is efficiently parameterized using a scaling vector $\mathbf{s}_i \in \mathbb{R}^3$ and a rotation quaternion $\mathbf{q}_i \in \mathbb{R}^4$, such that $\boldsymbol{\Sigma}_i = \mathbf{R}_i \mathbf{S}_i \mathbf{S}_i^{T} \mathbf{R}_i^{T}$, where $\mathbf{R}_i$ is the rotation matrix derived from $\mathbf{q}_i$ and $\mathbf{S}_i$ is a diagonal scaling matrix derived from $\mathbf{s}_i$.
To render an image, these 3D Gaussians are projected onto the 2D image plane and then ``splatted''. The final pixel color $\mathbf{C}$ is computed by alpha-blending the contributions from $N$ Gaussians sorted by depth:
\begin{equation}
    \mathbf{C} = \sum_{i=1}^{N} \mathbf{c}_i \alpha_i \prod_{j=1}^{i-1}(1-\alpha_j)\ ,
\end{equation}
where $\alpha_i$ here denotes the projected 2D Gaussian's opacity, derived from $\alpha_i$ and the 2D covariance. This differentiable rendering process allows for end-to-end optimization of all Gaussian parameters ($\boldsymbol{\mu}_i, \mathbf{s}_i, \mathbf{q}_i, \alpha_i, \mathbf{c}_i$) using gradient descent, typically initialized from a sparse point cloud obtained via Structure-from-Motion (SfM). 3DGS employs adaptive density control, involving splitting, cloning, and pruning Gaussians during optimization. 

\noindent\textbf{4D Gaussian Splatting.}
4D Gaussian Splatting (4DGS)~\cite{4dgs} adapts 3DGS for dynamic scenes. We build upon Ex4DGS~\cite{ex4dgs}, which separates Gaussians into static ($\mathcal{G}_s$) and dynamic ($\mathcal{G}_d$) components. Static Gaussians ($\mathcal{G}_s$) maintain constant covariance ($\boldsymbol{\Sigma}_s$), opacity ($\alpha_s$), and SH coefficients ($\mathbf{c}_s$), with only position $\boldsymbol{\mu}_s(t)$ changing linearly over the scene duration $L$:
\begin{equation}
\boldsymbol{\mu}_s(t) = \boldsymbol{\mu}_0 + \frac{t}{L} \boldsymbol{\mu}_{\text{disp}}
\end{equation}
where $\boldsymbol{\mu}_0$ is a pivot position and $\boldsymbol{\mu}_{\text{disp}}$ is the learned translation. Dynamic Gaussians ($\mathcal{G}_d$) use a keyframe approach, storing parameters only at sparse, uniformly sampled keyframes $\mathcal{K} = \{t \mid t = nI, n \in \mathbb{Z}, t \in \mathcal{T}\}$, where $\mathcal{T}$ represents the set of all timestamps in the entire video sequence. Properties at intermediate times $t$ are interpolated: position $\boldsymbol{\mu}_d(t)$ using Cubic Hermite Interpolation (CHip), quaternion rotation $\mathbf{q}_d(t)$ using Spherical Linear Interpolation (Slerp), and temporal opacity $\sigma_t(t)$ using a custom two-Gaussian mixture model (parameterized by means $a_s^o, a_f^o$ and variances $b_s^o, b_f^o$) to handle appearance/disappearance. For dynamic Gaussians in Ex4DGS, scale ($\mathbf{s}_d$) and SH coefficients ($\mathbf{c}_d$) are typically assumed constant over time.

\subsection{4D Dynamic Gaussian Compression}
\label{sec:4d_dynamic_compression}

Building upon the Ex4DGS framework~\cite{ex4dgs}, we introduce an end-to-end compression framework specifically tailored for dynamic 4D Gaussian representations. We follow the general principles established by Wang et al.~\cite{wang2024rdo3dgs} for compatibility with 3DGS compression, extending them to accommodate the unique challenges presented by dynamic scenes.

\noindent\textbf{Gaussian and SH Coefficient Masking.}
To reduce model complexity, redundant Gaussians are pruned using learnable masks $\phi_i$. A soft mask $\phi_{i}^{\text{soft}} = \text{sigmoid}(\phi_{i})$ is generated, then binarized to a hard mask $\phi_{i}^{\text{hard}} \in \{0, 1\}$ using a threshold $\phi_{\text{thres}}$ and the straight-through estimator (STE): $\phi_{i}^{\text{hard}} = \text{sg}(\mathbb{I}(\phi_{i}^{\text{soft}} > \phi_{\text{thres}}) - \phi_{i}^{\text{soft}}) + \phi_{i}^{\text{soft}}$. Here, $\text{sg}(\cdot)$ denotes the stop-gradient operation, and $\mathbb{I}(\cdot)$ is the indicator function. This hard mask is applied to key attributes like scale $s_i$ and opacity $\alpha_i$. The associated pruning cost, contributing to the rate term $\mathcal{L}_{\text{rate}}$ (the full formulation of which is detailed in \Cref{sec:appendix_rate_loss}), is:
\begin{equation}
\mathcal{L}_{\text{GSprune}} = \frac{1}{N} \sum_{i} \phi_{i}^{\text{soft}}
\end{equation}
Similarly, redundant SH coefficients are adaptively pruned using learnable masks $\theta_i^{(l)}$ per Gaussian $i$ and degree $l \ge 1$. Soft masks $\theta_{i}^{(l)\text{soft}} = \text{sigmoid}(\theta_{i}^{(l)})$ are binarized to hard masks $\theta_{i}^{(l)\text{hard}}$ via STE using a threshold $\theta_{th}$, too.
The hard mask is applied to zero out corresponding coefficients $\hat{c}_{i}^{(l)} = \theta^{(l)\text{hard}} c_{i}^{(l)}$. The weighted pruning loss contributing to $\mathcal{L}_{\text{rate}}$ is:
\begin{equation}
\mathcal{L}_{\text{SHprune}} = \frac{1}{N} \sum_{i} \sum_{l=1}^{k} \frac{2l+1}{(k+1)^2 - 1} \theta_{i}^{(l)\text{soft}},
\end{equation}

where $k$ is the maximum degree of the SHs, typically set to 3. This masking strategy is uniformly applied to both static and dynamic Gaussians, and experiments confirmed effective pruning across both categories. The results are available in \Cref{appendix:compression_ratio}.

\noindent\textbf{SH Coefficient, Rotation, and Scale Compression.}
Spherical Harmonics (SH) coefficients, rotation, and scale parameters exist for both static and dynamic Gaussians. These parameters are compressed using entropy-constrained vector quantization (ECVQ)~\cite{ecvq}, following Wang et al.~\cite{wang2024rdo3dgs}. Recognizing potential differences in compression sensitivity, scalar quantization is selectively applied.

Specifically, rotation parameters for dynamic Gaussians differ from static Gaussians in that they are defined at each timestamp. Although storing residual rotations was initially considered, preliminary experiments indicated that the entropy gain was marginal compared to the complexity of maintaining additional codebooks. Hence, we opted to store all rotation parameters directly.

\noindent\textbf{Opacity.}
Unlike traditional 3D Gaussian methods, Ex4DGS parameterizes dynamic opacity temporally by defining a scale function with center parameters ($a_s^o, a_f^o$) and variance parameters ($b_s^o, b_f^o$), effectively acting as an on/off temporal mask multiplied with a base opacity ($\alpha^d$). Given that the base opacity for dynamic points shares similar properties with the static opacity ($\alpha^s$), we follow Wang et al.~\cite{wang2024rdo3dgs} and apply scalar quantization to the base opacity. However, the temporal center and variance parameters, being expressed in time units rather than opacity values, exhibit different sensitivities to quantization. Empirical tests indicated that variance parameters ($b_s^o, b_f^o$) were particularly sensitive to quantization artifacts, hence, we avoid aggressive quantization for these parameters to maintain visual fidelity. Detailed empirical validation of this decision is presented in \Cref{subsec:ablation}.

\subsection{Wavelet Compression for Dynamic Positions}
To efficiently represent the motion trajectories of dynamic Gaussians ($\boldsymbol{\mu}_d(t) \in \mathbb{R}^{T\times3}$), we employ a single-level Haar wavelet transform along the temporal dimension. The wavelet transform is specifically chosen for its excellent time-frequency localization, enabling an efficient representation of motion trajectories that are generally smooth but may contain localized, high-frequency details (e.g., sharp turns or accelerations). Formally, given the dynamic positions $\boldsymbol{\mu}_d = [\mathbf{p}_1, \mathbf{p}_2, \dots, \mathbf{p}_T]^\top$ with $\mathbf{p}_t \in \mathbb{R}^3$, the wavelet transform is applied as:

\begin{equation}
\mathbf{F} = \mathbf{W}\boldsymbol{\mu}_d, \quad \mathbf{F} =
\begin{bmatrix}
\mathbf{F}_a \
\mathbf{F}_d
\end{bmatrix}
\end{equation}

Here, $\mathbf{W}$ denotes the Haar wavelet transform matrix, decomposing the trajectory into approximation (low-frequency) coefficients $\mathbf{F}_a \in \mathbb{R}^{\frac{T}{2}\times3}$ and detail (high-frequency) coefficients $\mathbf{F}_d \in \mathbb{R}^{\frac{T}{2}\times3}$. Recognizing that coarse motion patterns captured by $\mathbf{F}_a$ are perceptually more significant, we explicitly discard the detail coefficients $\mathbf{F}_d$ by zeroing them out. And the compressed trajectory $\hat{\boldsymbol{\mu}}_d$ is then reconstructed via the inverse Haar transform, which is equivalent to multiplying by the transpose of $\mathbf{W}$ due to its orthogonality:

\begin{equation}
\mathbf{F}_{\text{masked}} =
\begin{bmatrix}
\mathbf{F}_a \
\mathbf{0}
\end{bmatrix}, \ \ \hat{\boldsymbol{\mu}}_d = \mathbf{W}^\top \mathbf{F}_{\text{masked}}
\end{equation}

This approach significantly reduces storage and transmission requirements while effectively preserving essential dynamic motion information.

\subsection{End-to-End Rate-Distortion Optimization}
Following Bell\'e et al.~\cite{balle2017end}, our framework performs end-to-end rate-distortion optimization. Our framework jointly optimizes all relevant parameters by minimizing a composite loss function. These parameters include the original Ex4DGS parameters, those from our differentiated quantization (such as codebooks and entropy models for ECVQ), and the wavelet-compressed dynamic position representations. This total loss, $\mathcal{L}_{\text{total}}$, integrates the standard rate-distortion objective with regularization terms inherent to Ex4DGS:

\begin{equation}
\mathcal{L}_{\text{total}} = \mathcal{L}_{\text{dist}} + \lambda_{\text{R}} \mathcal{L}_{\text{rate}} + \lambda_{\text{reg}} \mathcal{L}_{\text{reg}}
\end{equation}

Here, $\mathcal{L}_{\text{dist}}$ represents the reconstruction fidelity loss, typically combining L1 loss and D-SSIM between the rendered image $\mathbf{C}$ and the ground truth image $\mathbf{C}_{\text{GT}}$:
\begin{equation}
\mathcal{L}_{\text{dist}} = (1.0 - \lambda_{\text{dssim}}) \mathcal{L}_{L1}(\mathbf{C}, \mathbf{C}_{\text{GT}}) + \lambda_{\text{dssim}} (1.0 - \text{SSIM}(\mathbf{C}, \mathbf{C}_{\text{GT}}))
\end{equation}

The term $\mathcal{L}_{\text{rate}}$ quantifies the estimated bitrate cost of the compressed representation. It aggregates several components aimed at reducing model size. These include entropy-based costs derived from quantized attribute representations (\textit{e.g.} VQ indices) and potentially auxiliary losses that promote sparsity within the model, such as terms encouraging the pruning of redundant Spherical Harmonics coefficients or entire Gaussian primitives. These components are collectively balanced by the rate weighting factor $\lambda_{\text{R}}$. When ECVQ is employed, the objective intrinsically balances the rate cost, and the quantization distortion for the respective attributes during codeword selection and codebook learning. Furthermore, $\mathcal{L}_{\text{reg}}$ incorporates regularization terms adapted from Ex4DGS~\cite{ex4dgs} to encourage plausible scene dynamics. This includes terms penalizing excessive linear displacement for static points and terms ensuring temporal smoothness in the motion and rotation trajectories defined by the dynamic keyframes.

By optimizing this comprehensive loss function end-to-end, our framework learns the compression mechanisms (quantization, pruning, wavelet processing via its impact on $\mathcal{L}_{\text{dist}}$ and $\mathcal{L}_{\text{rate}}$) concurrently with the scene representation itself. This joint process directly balances the bitrate costs against the final rendering quality and dynamic consistency, guided by the various lambda hyperparameters.

\section{Experiment}
\label{sec:experiment}

\subsection{Experimental Settings}
\noindent\textbf{Datasets.} To validate the effectiveness of the proposed method, we conduct various experiments on two real-world datasets, including Neural 3D Video (N3V)~\cite{li2022n3v} and Technicolor~\cite{sabater2017techni}. Because the proposed method is built upon Ex4DGS~\cite{ex4dgs}, we match the evaluation protocol, it is originally adopted by the previous work~\cite{attal2023hyperreel}. Specifically, we use the entire set of N3V and the five scenes of Technicolor (Birthday, Fabien, Painter, Theater, and Train).

\noindent\textbf{Metrics.} As in the previous works, we report standard metrics including Peak Signal-to-Noise Ratio (PSNR), Structural Similarity Index Measure (SSIM), and Learned Perceptual Image Patch Similarity (LPIPS)~\cite{lpips} for quantitative evaluation. We also measure the storage size for each compression level to evaluate if the proposed method achieves reasonable performance across different compression strengths. By varying the compression parameters, we generate rate-distortion (RD) curves to comprehensively analyze the trade-off between visual quality (distortion) and the storage size (rate) of a single model that includes the entire video sequence. Additionally, we report frames-per-second (FPS) to measure the rendering speed.

\noindent\textbf{Implementation Details.} We first train the base model using Ex4DGS~\cite{ex4dgs} until densification terminates. Subsequently, we apply our proposed RD-optimized method, incorporating Gaussian point pruning, adaptive spherical harmonics (SH) pruning, entropy-constrained vector quantization (ECVQ), and the novel pointwise wavelet transform. To systematically analyze performance at varying compression strengths, we define six compression levels (Levels 1–6), adjusting pruning hyperparameters as follows: for Gaussian point pruning, $\lambda_{\text{GSprune}}$ is set to [0.05, 0.02, 0.01, 0.005, 0.002, 0.0005]; for SH coefficient pruning, $\lambda_{\text{SHprune}}$ is set to [0.5, 0.2, 0.1, 0.05, 0.02, 0.005], respectively. All experiments are conducted on an NVIDIA RTX 3090 GPU. On average, with this GPU setup, the initial Ex4DGS pre-training takes approximately 1 hour, and our subsequent ECVQ and pruning optimization stage takes another hour, totaling around 2 hours of training time.

\figntvRD
\figtechniRD
\figntvablation
\figtechniQual
\fignthreevQual

\subsection{Rate-Distortion Performance}
In this section, we analyze the end-to-end rate-distortion (RD) optimized compression performance of the proposed method. 
Our primary goal is to leverage the spatio-temporal characteristics of videos to achieve compression efficiency beyond the limits of existing methods, while offering flexible control over the rate-distortion trade-off. This goes beyond simple masking and quantization by incorporating video-specific properties. To achieve different operating points on the RD curve, we generate compressed Gaussians of varying sizes by adjusting the strength of Spherical Harmonics (SH) and Gaussian point pruning and quantization, similar to RD3DGS~\cite{wang2024rdo3dgs}. This is primarily controlled by adjusting the weight $\lambda$ for $\mathcal{L}_{\text{GSprune}}$ and $\mathcal{L}_{\text{SHprune}}$. \Cref{fig:rdo_n3v,fig:rdo_techni} present the rate-distortion curves (PSNR, SSIM, LPIPS vs. Model size in MB) for the N3V dataset~\cite{li2022n3v} and the Technicolor dataset~\cite{sabater2017techni}, respectively. 
Due to the space constraint, we provide the per-scene results in \Cref{appendix:all_tab}. 
Furthermore, \Cref{fig:qual_n3v,fig:qual_techni} provide qualitative 
\tabAblation
comparisons, showcasing the visual fidelity of our method at key compression levels against original Ex4DGS~\cite{ex4dgs}.
The results demonstrate that our proposed approach, incorporating techniques like differentiated quantization and the wavelet transform, achieves significant model size reduction (e.g., up to 91$\times$ compression) compared to Ex4DGS while maintaining reasonable visual quality. This validates the effectiveness of our method in providing a flexible trade-off between compression rate and rendering quality suitable for diverse computational requirements.

\subsection{Ablation Study}
\label{subsec:ablation}

\noindent\textbf{Opacity Quantization.} We validate the opacity quantization strategy described in \Cref{sec:4d_dynamic_compression} through ablation experiments summarized in \Cref{tab:ablation}. Recognizing that each parameter contributes differently to the final visual quality, we designed this experiment to identify the optimal subset of parameters for quantization that best balances compression efficiency and rendering fidelity, rather than naively quantizing all components. Specifically, we examine the quantization sensitivity of four parameters: static opacity ($\alpha^s$), dynamic opacity ($\alpha^d$), center parameters ($a_s^o, a_f^o$), and variance parameters ($b_s^o, b_f^o$). 
Although static and dynamic opacity share similar characteristics, we evaluate them separately, considering the potential heightened sensitivity of dynamic opacity.
The results demonstrate that quantizing additional opacity-related parameters consistently improves storage efficiency. However, notably in the last two rows, quantizing variance parameters significantly reduces PSNR despite achieving marginal additional storage benefits. Based on the observation, we conclude that quantization should be limited to the static and dynamic opacity and center parameters to maintain a balanced trade-off between rendering fidelity and compression efficiency.

\noindent\textbf{Wavelet Transform for Dynamic Positions.} We further evaluate the effectiveness of wavelet-based compression applied specifically to the dynamic position trajectories ($\boldsymbol{\mu}_d(t)$). As detailed in \Cref{sec:method}, our proposed method leverages a Haar wavelet transform to capture low-frequency motion components while explicitly discarding high-frequency details. \Cref{fig:ablation} compares the rate-distortion performance of our method with wavelet transform against the baseline without it, clearly highlighting the benefit of incorporating wavelet compression. As shown in the figure, the rate-distortion curve of our method ('Ours') is consistently positioned to the top-left of the baseline ('w/o Wavelet Transform'), indicating a superior performance trade-off. This positioning signifies that for any given model size, our wavelet-based approach yields a higher PSNR, and conversely, for any target PSNR, it requires a smaller model size. Notably, employing the wavelet transform consistently improves performance across all tested compression levels, not only reducing the storage size but also enhancing rendering fidelity. Specifically, at compression level 1, our method achieves approximately 19\% storage reduction alongside a PSNR gain of 0.19 dB. Even at the highest fidelity setting (level 6), we observe around 26\% storage savings coupled with a PSNR improvement of 0.09 dB. These results emphasize that exploiting the inherent smoothness of point trajectories through wavelet-based compression not only effectively reduces redundancy in dynamic positions but also enhances trajectory modeling accuracy, leading to improved overall RD performance.

\noindent\textbf{Multi-Level Wavelet Decomposition.} To further analyze the impact of our wavelet compression, we conduct an ablation study on the level of wavelet decomposition applied to dynamic position trajectories. In addition to our proposed single-level Haar wavelet transform (retaining 1/2 of the coefficients),
\tabMultiLevelWavelet
we test more aggressive 2-level (retaining 1/4) and 3-level (retaining 1/8) decompositions. As summarized in \Cref{tab:ablation_wavelet_multilevel}, applying a deeper multi-level decomposition progressively reduces the model size. For instance, at compression Level 1, the size decreases from 1.63 MB to 1.16 MB. However, this gain in compression comes at a consistent cost to rendering quality. Crucially, the new operating points generated by the multi-level transforms do not surpass the RD curve established by our original single-level method.

\noindent\textbf{Analysis on Higher-Fidelity Trade-offs.} Our framework's compression is achieved through multiple lossy mechanisms, including pruning, ECVQ, and the wavelet transform. Therefore, we explore whether adjusting parameters beyond pruning can yield higher-quality results. Specifically, starting from our Level 6 setting, we test two strategies: (1) reducing the rate-loss weight, $\lambda_R$, in our objective function, and (2) additionally increasing the ECVQ codebook size.
\setcounter{footnote}{0}\tabHigherFid
The results, averaged on the N3V dataset, are presented in \Cref{tab:ablation_high_fidelity}. Reducing $\lambda_R$ does not affect the performance and size significantly. In addition, enlarging the codebook provides a minor additional PSNR gain, it nearly doubles the model size, indicating a less efficient trade-off. This study demonstrates that the future direction can be to find a new way to control or to push the curve to the better front.


\footnotetext{The PSNR and Size values for our Level 1 and 6 models differ from those in \Cref{tbl:n3v_compare} as a different pre-trained model was used for this ablation study. For a detailed explanation, please see \Cref{appendix:note_on_pretraining}.}

\subsection{Comparative Evaluation}

\tabNthreeV

This section evaluates the proposed method against existing models. Our primary goal is a framework for flexible rate-distortion trade-offs in dynamic 4D scene representation, not just state-of-the-art rendering. The comparison highlights the method's quality-size characteristics and its ability to significantly compress data while maintaining reasonable fidelity.

We report the PSNR, storage size, and FPS for the N3V dataset in \Cref{tbl:n3v_compare}. The most notable advantage of our method is its significant size reduction. At compression Level 1, our model achieves a remarkable 98.9\% size reduction compared to the baseline Ex4DGS, resulting in an average size of only 1.26 MB. Even when compared to 4DGaussians, the previously most storage-efficient method, our approach demonstrates over 25 times greater efficiency at the same compression level. 
Our method significantly outperforms existing models in rendering speed, reaching up to 163 FPS at Level 1, suitable for real-time applications. While there is an expected trade-off in rendering quality (PSNR) due to aggressive compression,
\tabTechni
our model still produces results that are sufficiently competitive. Particularly noteworthy is the comparison at Level 6, where our method not only has a smaller model size (11.06 MB) but also a higher average PSNR (29.66 dB) compared to 4DGaussians (34 MB, 28.63 dB). A similar trend is observed for the Technicolor dataset, as summarized in \Cref{tbl:techni_compare}.

\vspace{-5pt}
\section{Discussion}
\vspace{-5pt}
\noindent\textbf{Conclusion.} We presented the first end-to-end rate-distortion (RD) optimized compression framework for 4D Gaussian Splatting (4DGS) to address the challenge of large model sizes hindering deployment. Leveraging Ex4DGS, our method uses static/dynamic decomposition with adaptive RD-guided quantization and pointwise wavelet transform to achieve flexible compression. Although this work builds upon the Ex4DGS framework, the core principles of our rate-distortion optimized compression are general and can be adapted to other 4DGS models. Our approach demonstrates significant model size reduction (down to approximately 1.1\% of the original size ) while maintaining reasonable visual quality across standard benchmarks and offering user control over the rate-quality trade-off. This significantly improves the practicality of 4DGS for real-time rendering on diverse computational platforms, including resource-constrained devices. 

\noindent\textbf{Limitation.} Our approach of filtering high-frequency motion via a wavelet transform has an inherent limitation in handling excessively dynamic movements, which can manifest as motion blur artifacts on fast-moving objects, particularly at high compression levels. Additionally, dynamic points still account for a substantial portion of total storage, highlighting significant room for improvement in compressing dynamic components (\Cref{appendix:vis_static_dynamic,appendix:sizes}). Future work includes refining the loss function, for example by applying separate masking weights for static and dynamic points, to further enhance dynamic point compression and overall visual quality.

\noindent\textbf{Broader Impacts.} Our work significantly simplifies the storage, transmission, and deployment of volumetric videos, directly contributing to the widespread adoption and accessibility of volumetric content. In the long term, we anticipate that this research can serve as a foundational step toward establishing standards for volumetric video formats. Nevertheless, easier distribution may also enable rapid proliferation of malicious or inappropriate volumetric content, underscoring the need for proactive measures to mitigate such risks.




{
\small
\bibliographystyle{unsrt}
\bibliography{ref}
}


\appendix
\newpage

\section{Detailed Rate Loss Formulation}
\label{sec:appendix_rate_loss}

In \Cref{sec:4d_dynamic_compression}, we introduced the rate term $\mathcal{L}_{\text{rate}}$, which is a core component of our end-to-end rate-distortion optimization framework. The structure of our rate loss is based on the formulation proposed for static scenes in \cite{wang2024rdo3dgs}, which we adapt to the unique challenges of dynamic 4D Gaussian Splatting by extending it to handle both static and dynamic components separately.

This section provides the detailed formulation of $\mathcal{L}_{\text{rate}}$. The total rate loss is a weighted sum of four distinct terms: Gaussian pruning loss ($\mathcal{L}_{\text{GSprune}}$), SH coefficient pruning loss ($\mathcal{L}_{\text{SHprune}}$), an entropy loss ($\mathcal{L}_{\text{entropy}}$), and a vector quantization loss ($\mathcal{L}_{\text{VQ}}$).

The complete expression for the total rate loss is:
\begin{equation}
\mathcal{L}_{\text{rate}} = \lambda_{\text{GSprune}}\mathcal{L}_{\text{GSprune}} + \lambda_{\text{SHprune}}\mathcal{L}_{\text{SHprune}} + \mathcal{L}_{\text{entropy}} + \mathcal{L}_{\text{VQ}}
\end{equation}

Here, $\mathcal{L}_{\text{GSprune}}$ and $\mathcal{L}_{\text{SHprune}}$ are the pruning losses defined in Equations (4) and (5) of the main text, which encourage model sparsity. The terms $\mathcal{L}_{\text{entropy}}$ and $\mathcal{L}_{\text{VQ}}$ arise from the entropy-constrained vector quantization (ECVQ) applied to various Gaussian attributes (e.g., scale, rotation, SH coefficients).

The entropy loss, $\mathcal{L}_{\text{entropy}}$, estimates the bitrate required to encode the quantized parameters. It is formulated as the cross-entropy between the distribution of the quantized symbols and our learned entropy model:
\begin{equation}
\small
\mathcal{L}_{\text{entropy}} = \frac{1}{N} \sum_{m \in \{\text{static, dynamic}\}} \sum_{i=1}^{N} \left( \frac{r_{i,j}^{(s,m)}}{\lambda^{(s,m)}} + \frac{r_{i,j}^{(r,m)}}{\lambda^{(r,m)}} + \frac{r_{i,j}^{(DC,m)}}{\lambda^{(DC,m)}} + \frac{r_{i,j}^{(SH1,m)}}{\lambda^{(SH1,m)}} + \frac{r_{i,j}^{(SH2,m)}}{\lambda^{(SH2,m)}} + \frac{r_{i,j}^{(SH3,m)}}{\lambda^{(SH3,m)}} \right)
\end{equation}

The vector quantization loss, $\mathcal{L}_{\text{VQ}}$, measures the distortion introduced by mapping continuous parameter values to discrete codebook vectors during quantization:
\begin{equation}
\mathcal{L}_{\text{VQ}} = \frac{1}{N} \sum_{m \in \{\text{static, dynamic}\}} \sum_{i=1}^{N} \left( d_{i,j}^{(s,m)} + d_{i,j}^{(r,m)} + d_{i,j}^{(DC,m)} + d_{i,j}^{(SH1,m)} + d_{i,j}^{(SH2,m)} + d_{i,j}^{(SH3,m)} \right)
\end{equation}

In these equations, the superscripts $(s,m), (r,m), \dots$ denote different attributes such as scale and rotation for both static and dynamic ($m$) Gaussians. The term $r_{i,j}$ represents the estimated bit cost (rate), while $d_{i,j}$ signifies the quantization error (distortion).

\section{Additional Quantitative Evaluation Results}
\label{appendix:quantitative}
We report the detailed quantitative evaluation results, including storage size and rendering fidelity metrics (PSNR, SSIM, LPIPS), for each scene of the N3V and Technicolor datasets in \Cref{appendix:all_tab}. The most notable highlight is the substantial model size reduction achieved by our proposed method across all scenes, ranging from 79.3\% at the highest fidelity setting (Level 6) to 99.3\% at the most aggressive compression level (Level 1). Although aggressive compression (lower levels) occasionally leads to significant drops in rendering fidelity, our framework provides multiple trade-off options to flexibly adapt according to deployment scenarios and resource constraints. These results indicate promising directions for future work, focusing on enhancing reconstruction fidelity at lower compression levels to further broaden applicability.

\tabAllAppendix

\newpage
\section{Rate-Distortion Performance for Each Scene}
\label{appendix:rd_performance}
In addition to the numerical results provided in \Cref{appendix:quantitative}, we illustrate detailed per-scene rate-distortion (RD) curves for both datasets. Specifically, \Cref{fig:rdo_n3v_perscene,fig:rdo_techni_perscene} show RD curves for the N3V and Technicolor datasets, respectively. These figures visually reaffirm the notable size reductions achieved by our method, consistently demonstrating drastic reductions in model sizes. A closer examination reveals that fidelity degradation for the Technicolor dataset is relatively moderate compared to N3V, likely due to the simpler and less intricate motions inherent in Technicolor scenes. Conversely, N3V contains complex and detailed actions, such as precise tool movements or dynamic interactions (e.g., operating kitchen tongs or using a torch to sear meat), making compression-induced fidelity degradation more pronounced. This observation highlights opportunities for further enhancing compression methods, specifically targeting more effective exploitation of detailed volumetric motion information. Another noteworthy finding appears in the Technicolor dataset’s “Train” scene, where certain compression levels achieve superior rendering fidelity compared to the baseline Ex4DGS method. This result indicates the potential benefit of model simplicity, characterized by fewer points and lower entropy, which can unexpectedly enhance view-synthesis quality by effectively reducing model redundancy.
\figntvRDEach
\figtechniRDEach

\clearpage

\section{Analysis on the Number of Pruned Points}
\label{appendix:compression_ratio}

To verify that the proposed method effectively reduces redundancy in both dynamic ($\mathcal{G}_d$) and static ($\mathcal{G}_s$) components, we examine the reduction ratios across different compression levels, as reported in \Cref{tab:ratio}. The results indicate that at the highest compression setting (Level 1), the reduction rates for dynamic and static points are similarly significant. However, at the lowest compression level (Level 6), the reduction ratio of dynamic points consistently surpasses that of static points. This suggests that when fidelity is prioritized, our method tends to prune dynamic points more aggressively, likely due to the smaller spatial coverage of dynamic points compared to static points. Nonetheless, given that dynamic points typically correspond to salient foreground elements, overly aggressive pruning of dynamic points may adversely impact overall visual quality. This highlights an avenue for future improvement: dynamically adjusting the pruning strategy—for example, by assigning separate weighting coefficients to static and dynamic points within the pruning loss—to better preserve visually salient dynamic components while maintaining high compression efficiency.
\tabCompRatioSD

\section{Comparison with Concurrent Work: Light4GS}
\label{appendix:light4gs_comparison}

This section provides a direct comparison with the concurrent work Light4GS \cite{liu2025light4gs}. To ensure a fair comparison, we retrained our model to match the experimental protocol of Light4GS, which uses a different resolution ($1024\times768$) and a slightly different subset of the N3V dataset. The results are summarized in \Cref{tab:appendix_light4gs}.

The comparison highlights the different strengths of the two methods. While Light4GS achieves a higher absolute PSNR, our method demonstrates superior performance in terms of compression ratio and rendering speed, particularly in the high-compression setting. Specifically, our high-compression model is more than 2.5x smaller (1.48~MB vs. 3.77~MB) and runs approximately 4x faster ($\sim$160~FPS vs. $\sim$40~FPS).

Furthermore, the PSNR gap is not a fundamental limitation of our compression framework. As demonstrated in our ablation study on higher-fidelity trade-offs (\Cref{tab:ablation_high_fidelity}), relaxing the rate-loss weight ($\lambda_R$) can significantly boost our model's PSNR. This suggests the performance difference also stems from different design priorities and baseline models (Ex4DGS for ours, 4DGS for Light4GS), rather than the compression methodology itself. Our framework's primary advantage lies in its flexibility, offering a wide, user-controllable spectrum of rate-distortion-speed trade-offs suitable for diverse deployment scenarios.
\tabLightFourGS

\section{Note on Pre-trained Models for Ablation Studies}
\label{appendix:note_on_pretraining}

Our RD-optimized framework follows a two-stage training process: initial pre-training using the Ex4DGS methodology, followed by RD-optimized fine-tuning. For consistency, all primary experimental results reported in this paper (except for \Cref{tab:ablation_wavelet_multilevel,tab:ablation_high_fidelity}) originate from a single, fixed pre-trained model.

The ablation studies presented in \Cref{tab:ablation_wavelet_multilevel,tab:ablation_high_fidelity} were conducted as supplementary experiments. As the original pre-trained model was unavailable at the time of these additional experiments, they utilized a separately pre-trained model. Our methodology involves initiating the RD-optimized fine-tuning after the densification stage of the Ex4DGS schedule, which is prior to its full convergence. Consequently, minor performance variations in the starting point for fine-tuning can occur between different pre-training runs. This explains the slight differences in the baseline performance for the Level 1 and Level 6 configurations between these tables and others in the paper. However, this does not affect our main conclusion: once a pre-trained model is fixed, our framework consistently demonstrates a flexible and effective rate-distortion trade-off controlled by the fine-tuning parameters.

\newpage
\section{Visualization of Static and Dynamic Points}
\label{appendix:vis_static_dynamic}
We visualize the static and dynamic components separately across different compression levels to further investigate the impact of our proposed compression method, as shown in \Cref{fig:staticdynamic_nthreev_first,fig:staticdynamic_nthreev_second,fig:staticdynamic}. Consistent with the quantitative findings discussed in \Cref{appendix:compression_ratio}, our method tends to prune dynamic points more aggressively than static points, despite dynamic regions generally holding higher visual saliency. Particularly noticeable in the highly compressed results (Level 1) shown in \Cref{fig:staticdynamic_nthreev_first,fig:staticdynamic_nthreev_second}, rapidly moving objects often exhibit noticeable local blur due to aggressive dynamic point pruning. Nevertheless, it’s worth emphasizing that even at moderate compression (Level 6), our method successfully achieves substantial storage savings (approximately 90\% on average) while preserving rendering quality across most scenes without significant visual artifacts. This demonstrates our method’s practical balance between storage efficiency and visual fidelity.

\figStaticDynamicNThreeVFirst
\figStaticDynamicNThreeVSecond
\figStaticDynamicTechni

\clearpage
\newpage

\section{Analysis of Compressed Representation Size}
\label{appendix:sizes}
This appendix details the storage allocation (in MB and percentage) for individual components of our RD-optimized model across various scenes from the Technicolor~\Cref{appendix:storagetechni} and N3V~\Cref{appendix:storagen3v} datasets. Results are shown for Low Rate (Level 1) and High Rate (Level 6) settings. These tables provide a granular view of bit distribution for key components such as sorting indices for the masked SH parameters (sort idx), entropy-coded indexes (Indexes), masked wavelet coefficients $\mathbf{F}_{\text{masked}}$, dynamic position displacements ($\boldsymbol{\mu}_{\text{disp}}$), static position ($\boldsymbol{\mu}_0$), codebooks, logits, and opacity variance parameters ($\beta_*^o$). 

Across both datasets, it can be observed that components related to the explicit representation of Gaussians and their dynamic properties (e.g., $\mathbf{F}_{\text{masked}}$, Indexes, sort idx) generally constitute a significant portion of the bitstream, especially at higher rates where more detail is preserved. The relative proportions vary across scenes, reflecting differences in scene complexity and motion characteristics.

\tabStorageNtVApp
\tabStorageApp


\newpage
\clearpage
\section*{NeurIPS Paper Checklist}

\begin{enumerate}

\item {\bf Claims}
    \item[] Question: Do the main claims made in the abstract and introduction accurately reflect the paper's contributions and scope?
    \item[] Answer: \answerYes{} 
    \item[] Justification: We include the summarization of the contribution and the problem scope in the introduction section, too.

\item {\bf Limitations}
    \item[] Question: Does the paper discuss the limitations of the work performed by the authors?
    \item[] Answer: \answerYes{} 
    \item[] Justification: The limitation is included in the ``Discussion'' section.

\item {\bf Theory assumptions and proofs}
    \item[] Question: For each theoretical result, does the paper provide the full set of assumptions and a complete (and correct) proof?
    \item[] Answer: \answerNA{} 
    \item[] Justification: Even if we provide the theoretical ground of the wavelet transformation, we do not provide the assumption or proof because we introduce the previously well studied wavelet transformation into the context of the 4D Gaussian Splatting.

    \item {\bf Experimental result reproducibility}
    \item[] Question: Does the paper fully disclose all the information needed to reproduce the main experimental results of the paper to the extent that it affects the main claims and/or conclusions of the paper (regardless of whether the code and data are provided or not)?
    \item[] Answer: \answerYes{} 
    \item[] Justification: We elaborate on the proposed method as well as the baseline to enhance reproducibility.

\item {\bf Open access to data and code}
    \item[] Question: Does the paper provide open access to the data and code, with sufficient instructions to faithfully reproduce the main experimental results, as described in supplemental material?
    \item[] Answer: \answerYes{} 
    \item[] Justification:  We will make the source code to reproduce the experimental results public as well as the pre-trained model. We already include the source code as the supplementary material. For data, we do not introduce any datasets.

\item {\bf Experimental setting/details}
    \item[] Question: Does the paper specify all the training and test details (e.g., data splits, hyperparameters, how they were chosen, type of optimizer, etc.) necessary to understand the results?
    \item[] Answer: \answerYes{} 
    \item[] Justification: We provide the details of the training and test datasets, hyperparameters of the newly proposed method, while following the baseline. In addition, the source code contains the detailed configuration files.

\item {\bf Experiment statistical significance}
    \item[] Question: Does the paper report error bars suitably and correctly defined or other appropriate information about the statistical significance of the experiments?
    \item[] Answer: \answerNo{} 
    \item[] Justification: Because the computational cost to run multiple times for all the experiments is too expensive, however, we verified that the training variance is small with three datasets.

\item {\bf Experiments compute resources}
    \item[] Question: For each experiment, does the paper provide sufficient information on the computer resources (type of compute workers, memory, time of execution) needed to reproduce the experiments?
    \item[] Answer: \answerYes{} 
    \item[] Justification: We provide information related to the computational cost in the experiment section.
    
\item {\bf Code of ethics}
    \item[] Question: Does the research conducted in the paper conform, in every respect, with the NeurIPS Code of Ethics \url{https://neurips.cc/public/EthicsGuidelines}?
    \item[] Answer: \answerYes{} 
    \item[] Justification: We do not compose a new dataset and adopt only the well-discussed datasets.

\item {\bf Broader impacts}
    \item[] Question: Does the paper discuss both potential positive societal impacts and negative societal impacts of the work performed?
    \item[] Answer: \answerYes{} 
    \item[] Justification: We discuss the societal impacts in the discussion section.
    
\item {\bf Safeguards}
    \item[] Question: Does the paper describe safeguards that have been put in place for responsible release of data or models that have a high risk for misuse (e.g., pretrained language models, image generators, or scraped datasets)?
    \item[] Answer: \answerNA{} 
    \item[] Justification: The proposed method is related to 3D reconstruction of the multi-view videos. It does not have any risks of misuse.

\item {\bf Licenses for existing assets}
    \item[] Question: Are the creators or original owners of assets (e.g., code, data, models), used in the paper, properly credited and are the license and terms of use explicitly mentioned and properly respected?
    \item[] Answer: \answerYes{} 
    \item[] Justification: We maintain the original license.

\item {\bf New assets}
    \item[] Question: Are new assets introduced in the paper well documented and is the documentation provided alongside the assets?
    \item[] Answer: \answerNA{} 
    \item[] Justification: We do not introduce new assets.

\item {\bf Crowdsourcing and research with human subjects}
    \item[] Question: For crowdsourcing experiments and research with human subjects, does the paper include the full text of instructions given to participants and screenshots, if applicable, as well as details about compensation (if any)? 
    \item[] Answer: \answerNA{} 
    \item[] Justification: This work does not involve crowdsourcing nor research with human subjects.

\item {\bf Institutional review board (IRB) approvals or equivalent for research with human subjects}
    \item[] Question: Does the paper describe potential risks incurred by study participants, whether such risks were disclosed to the subjects, and whether Institutional Review Board (IRB) approvals (or an equivalent approval/review based on the requirements of your country or institution) were obtained?
    \item[] Answer: \answerNA{} 
    \item[] Justification: This paper does not involve crowdsourcing nor research with human subjects.

\item {\bf Declaration of LLM usage}
    \item[] Question: Does the paper describe the usage of LLMs if it is an important, original, or non-standard component of the core methods in this research? Note that if the LLM is used only for writing, editing, or formatting purposes and does not impact the core methodology, scientific rigorousness, or originality of the research, declaration is not required.
    \item[] Answer: \answerNA{} 
    \item[] Justification: We use LLMs only for improving the grammatical errors in writing.

\end{enumerate}


\begin{equation*}
\begin{split}
& \mathcal{L}_{\text{total}} = \mathcal{L}_{\text{dist}} + \lambda_{\text{R}} \mathcal{L}_{\text{rate}} + \lambda_{\text{reg}} \mathcal{L}_{\text{reg}}\\
&\mathcal{L}_{\text{VQ}} = \frac{1}{N} \sum_{m \in \{\text{static, dynamic}\}} \sum_{i=1}^{N} \left( d_{i,j}^{(s,m)} + d_{i,j}^{(r,m)} + d_{i,j}^{(DC,m)} + d_{i,j}^{(SH1,m)} + d_{i,j}^{(SH2,m)} + d_{i,j}^{(SH3,m)} \right)\\
&\mathcal{L}_{\text{entropy}} = \frac{1}{N} \sum_{m \in \{\text{static, dynamic}\}} \sum_{i=1}^{N} \left( \frac{r_{i,j}^{(s,m)}}{\lambda^{(s,m)}} + \frac{r_{i,j}^{(r,m)}}{\lambda^{(r,m)}} + \frac{r_{i,j}^{(DC,m)}}{\lambda^{(DC,m)}} + \frac{r_{i,j}^{(SH1,m)}}{\lambda^{(SH1,m)}} + \frac{r_{i,j}^{(SH2,m)}}{\lambda^{(SH2,m)}} + \frac{r_{i,j}^{(SH3,m)}}{\lambda^{(SH3,m)}} \right) 
\end{split}
\end{equation*}

\end{document}